\documentclass[aip,reprint]{revtex4-1}
\usepackage{amsmath}
\usepackage{amssymb}
\usepackage{units}
\usepackage{graphicx}
\usepackage{xcolor}
\usepackage{soul}

\draft 

\newcommand{\e}{\varepsilon}

\begin{document}


\title{Reinforcement learning for suppression of collective activity in oscillatory ensembles
} 

\author{Dmitrii Krylov}
\affiliation{Skolkovo Institute of Science and Technology, Bolshoy blvd. 30/1, Moscow, 121205, Russia}
\author{Dmitry V. Dylov}
\email{d.dylov@skoltech.ru}
\affiliation{Skolkovo Institute of Science and Technology, Bolshoy blvd. 30/1, Moscow, 121205, Russia}
\author{Michael Rosenblum}
\email{mros@uni-potsdam.de}
\affiliation{Institute of Physics and Astronomy, University of Potsdam,
Karl-Liebknecht-Str. 24/25, 14476 Potsdam-Golm, Germany}

\date{\today}

\begin{abstract}
We present a use of modern machine learning approaches to suppress self-sustained collective oscillations typically signaled by ensembles of degenerative neurons in the brain. The proposed hybrid model relies on two major components: an environment of oscillators and a policy-based reinforcement learning block. We report a model-agnostic synchrony control based on proximal policy optimization and two artificial neural networks in an Actor-Critic configuration. A class of physically meaningful reward functions enabling the suppression of collective oscillatory mode is proposed. The synchrony suppression is demonstrated for two models of neuronal populations -- for the ensembles of globally coupled limit-cycle Bonhoeffer-van der Pol oscillators and for the bursting Hindmarsh--Rose neurons using rectangular and charge-balanced stimuli.
\end{abstract}


\maketitle 

\begin{quotation}
Certain neurological diseases such as Parkinson's are believed to originate from the networks of degenerative neurons in the brain that collectively emit signals in an undesired synchronized way. Suppression of this synchronous dynamics has been an active direction of research on the edge of neuroscience, physics, and computer science in an attempt to improve efficacy of such widely accepted medical technique as deep brain stimulation (DBS). However, control of a large network of interacting neurons is a complicated nonlinear problem, which instigated a development of theoretical and numerical simulation approaches based on physical models that can mimic the behaviour of large neuronal ensembles. 

On the other hand, the explosive development of machine learning (ML) in recent years has offered a new data-driven methodology that could operate completely unaware of the physical world or of the underlying neuronal model. Here, an ML technique called reinforcement learning (RL) allowed us to control the undesired synchrony of the oscillatory ensembles. The signals were generated by popular physical models and then fed to train the ML module to undertake the data-driven control. Two artificial neural networks called Actor and Critic successfully learned how to suppress the collective mode, while relying on nothing but a reward for asynchrony. The suppression workflow proposed here is universal and could pave the way towards clinical realization of the DBS via reinforced learning.
\end{quotation}

\section{Introduction}
\label{intro}

Control of complex oscillatory networks is an important problem of nonlinear science,
with a number of practical applications. A particular challenge is to suppress collective 
activity that emerges due to synchronization in a population of interacting 
self-sustained units. This task is motivated by the ongoing research on efficient algorithms for the widely used medical technique called deep brain stimulation 
(DBS)~\cite{Benabid_et_al-91,*Benabid_et_al-09,*Kuehn-Volkmann-17}. 
DBS -- mostly used to treat Parkinson's disease -- entails high frequency open-loop 
pulse stimulation of certain brain regions via implanted micro-electrodes.
The stimulation is typically supplied around the clock and can significantly relieve the limb tremor associated with the disease.
The exact mechanisms of DBS are still a matter of debate~\cite{Johnson2008,*Gradinaru-09,*Deniau_et_al-10}, 
and the working hypothesis, exploited by the nonlinear science groups 
working in this field, assumes that the pathological brain activity emerges
due to an undesired synchronization of many interacting neurons. 
Correspondingly, it is hypothesized that the goal of the DBS is to desynchronize 
the large network of neurons, without suppressing the very oscillatory activity of 
its nodes~\cite{Tass-99}.
Theoretical studies and computer simulations suggest that this goal can 
be efficiently achieved by a feedback-based stimulation. 
Proposed schemes exploit delayed or non-delayed, linear and nonlinear 
control loops, with either continuous or pulsatile stimulation, and with adaptive tuning of the feedback 
parameters~\cite{Rosenblum-Pikovsky-04,*Rosenblum-Pikovsky-04a,%
Popovych-Hauptmann-Tass-05,%
Tukhlina-Rosenblum-Pikovsky-Kurths-07,Montaseri_et_al-13,Lin_2013,*Zhou_2017,Popovych_et_al-17}. 
An important advantage of these schemes is that they do not rely 
on the phase approximation, 
cf.~\cite{Tass-01,*Hauptmann-Tass-09,*Popovych-Tass-12,*Wilson-Moehlis-16,*Holt_et_al-16}.
Independently, the idea of feedback control is also being slowly developed  in the neuroscience 
community and some simple feedback algorithms have already been tested experimentally~\cite{Rosin-11,*Little-13}.

Another line of recent development is related to the rapid growth of machine 
learning (ML). The ML techniques are now extensively used for analysis and prediction of complex systems~\cite{Parlitz-18,*PhysRevLett.120.024102,*Zimmermann-Parlitz-18,%
*Quade-18,*Cestnik-Abel-19,*PhysRevE.99.042203,*Yeo-Melnyk-19}
and it is natural to try this framework for the purposes
of control as well. In this paper we propose to exploit an area of ML called reinforcement learning (RL) to suppress the collective mode in an oscillatory ensemble of globally coupled units.

Originally stemming from the optimal control problems and from the dynamical programming fields, reinforcement learning grew in popularity circa 1980~\cite{RL-Sutton1998}. RL algorithms differ from their predecessors in that they are capable of \emph{learning} the environment, effectively making it unnecessary to have a complete knowledge about the system in order to control it. 
There are two major angles to RL that have been actively implemented in various adjacent disciplines: a Q-function based approach~\cite{RL-surveymain-kaelbling1996reinforcement} and a policy gradient approach~\cite{RL-PPO-schulman2017}. The Q-function methods rely on a piece-wise or a discrete reward and are, therefore, sub-optimal for the task of continuous synchrony suppression problem at hand. Instead, we will consider the policy-based approach, with the policy $\pi$ implying the optimal strategy or a set of rules to emit stimuli signals at different times to suppress self-sustained oscillations. 

The policy gradient methods have many realizations based on stochastic gradient ascent algorithm~\cite{RL-Sutton1998}. In this work we will consider one of them, called Actor-Critic~\cite{RL-surveymain-kaelbling1996reinforcement}, and will conduct the Proximal Policy Optimization (PPO) for the purpose of synchrony suppression. Two artificial neural networks called Actor and Critic correspondingly are to be trained to suppress the collective mode, while relying on nothing but a reward for asynchrony. The role of the first network is to evaluate the policy, while the role of the other is to assess an advantage following each action by the Actor which sends the suppression stimuli to the environment. The overall Actor-Critic PPO RL module presented below can provide a robust data-driven control agnostic of the neuronal model and promises easy integration with current clinical DBS systems.  

\section{The model}

\begin{figure}
\centering
\includegraphics[width=1.1\columnwidth]{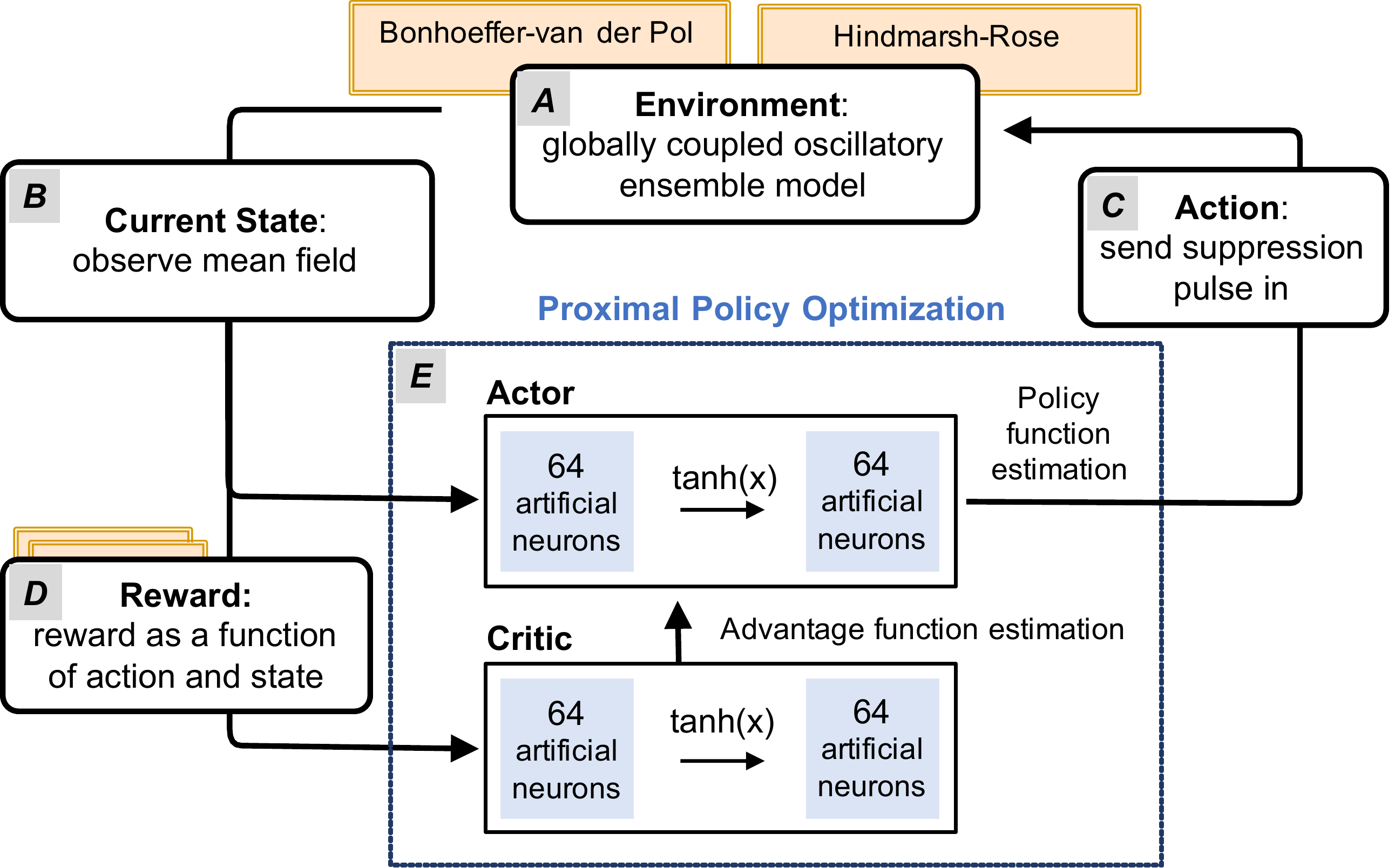}
\caption{Principle diagram of Reinforcement Learning via PPO Actor-Critic algorithm.}
\label{fig:diagram}
\end{figure}

In this work we consider a modern approach to RL based on proximal policy optimization with Actor and Critic algorithm~\cite{RL-stable-baselines} (see diagram of Fig.~\ref{fig:diagram}) to evaluate optimal policy $\pi$ for suppressing oscillations. There are five principal blocks that are involved in the control problem: (A)~Environment, (B)~Current State, (C)~Action block, (D)~Reward, (E)~PPO block containing two artificial neural networks (Actor and Critic). We will now describe each block and its function in detail.

\subsection{The Environment.}
The first key component of the diagram in Fig.~\ref{fig:diagram} is the environment that the RL algorithm needs to interact with in order to suppress synchrony. The ``environment'' could be either a real patient or a numerical model that simulates pathological dynamics of neuronal ensembles. Herein, we focus on the latter, considering two models of collective neuronal activity. In both cases we deal with globally coupled networks. The first type of network units are the periodic Bonhoeffer-van der Pol oscillators that mimic regularly spiking cells. The second type are the chaotically bursting Hindmarsh--Rose neurons.

\subsubsection{Bonhoeffer--van der Pol oscillators}

\begin{figure}
\centering
\includegraphics[width=0.75\columnwidth]{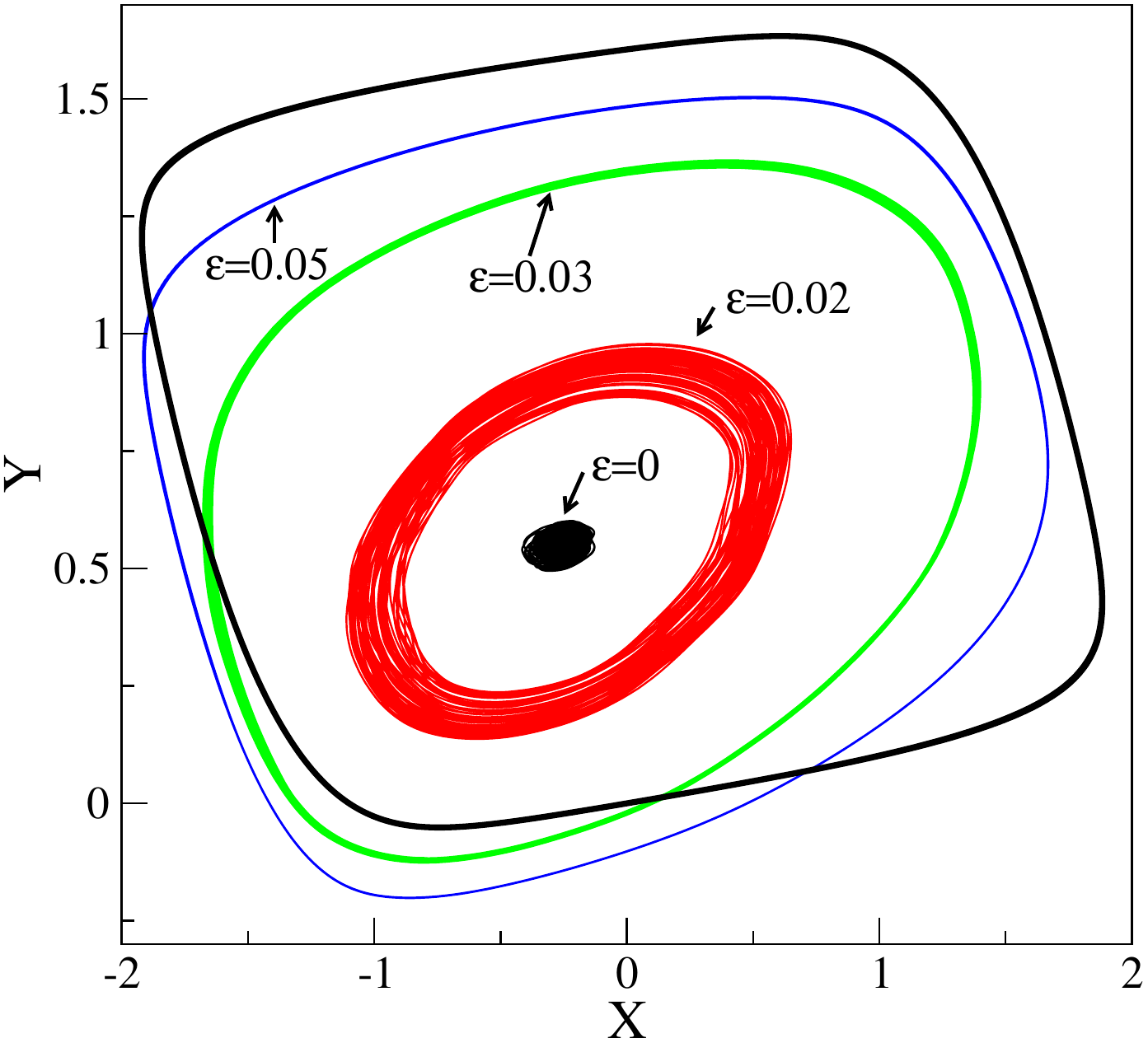}
\caption{Phase portraits for the collective mode, for $N=1000$ and for different values of
the coupling strength $\e$ of the ensemble of Bonhoefer--van der Pol oscillators, 
shown in mean-field coordinates $X=N^{-1}\sum_k x_k$, $Y=N^{-1}\sum_k y_k$. 
Notice that for $\e=0.02$ the collective dynamics is chaotic.
For comparison, a phase portrait of one unperturbed oscillator is shown by a thick line (the case $N=1$, $\e=0$ is shown).
}
\label{fig:BvdP1}
\end{figure}

Our first model is a population of $N$ globally coupled  Bonhoeffer--van der Pol oscillators:
\begin{equation}
\begin{cases}
\dot{x}_k &= x_k-x_k^3/3 - y_k +I_k +\e X + {\cal C}(t)\;,\\
\dot{y}_k &= 0.1 (x_k-0.8y_k+0.7)\;,\\
\end{cases}
\label{eq:bvdp}
\end{equation}
where $k=1,\ldots,N$, mean field $X=N^{-1}\sum_k x_k$, and ${\cal C}(t)$ is the action from the 
controller, specified below in Section~\ref{sec:action}.
The units are not identical: the currents $I_k$ are taken from a Gaussian distribution with the mean $0.6$ and standard deviation $0.1$. The strength of the global coupling is determined by the parameter $\e$.
Collective dynamics of this system is illustrated by the phase portraits shown in Fig.~\ref{fig:BvdP1}, 
where we plot $Y=N^{-1}\sum_k y_k$ vs. $X$ for different values of the coupling strength $\e$,
as well as the limit cycle of an isolated Bonhoeffer-van der Pol oscillator. 

This model has two properties that make the control problem non-trivial. 
First, for the sub-threshold coupling the  mean fields are $X_0\approx -0.27$, $Y_0\approx 0.55$,
i.e. the fixed point to be stabilized is not in the origin and is {\it a priori} unknown.
Next, this model exhibits chaotic collective dynamics for some values of $\e$ (see the broadened phase trajectory in Fig.~\ref{fig:BvdP1}). 

\subsubsection{Bursting Hindmarsh--Rose neuronal model}
The second type of oscillators that we consider is an ensemble of Hindmarsh-Rose \cite{Hindmarsh-Rose-84} neurons in a bursting regime:
\begin{equation}\label{eq:HRn}
\begin{cases}
 {\dot x}_k &= 3x_k^2 - x_k^3 + y_k - z_k + I_k +\e X+{\cal C}(t)\;,  \\ 
 {\dot y}_k &= 1 - 5x_k^2 - y_k \;, \\ 
 {\dot z}_k &= 0.006[4 (x_k +1.56 ) - z_k]\;. \\ 
 \end{cases}
\end{equation}
The currents $I_k$ are taken from a Gaussian distribution with the mean  
$3$ and the standard deviation $0.02$. For illustration, see Fig~\ref{fig:Bursting-TS} below.

\subsection{Current State Block}
\label{sec:state}
The Action and the Current State blocks in the algorithm shown in Fig.~\ref{fig:diagram} 
are merely the input \emph{to} and the resulting response output \emph{from} the Environment. 
We assume that both the input and the output are produced with a sampling rate $\delta$. 
In a real experiment, $\delta$ is determined by the hardware, namely by the analog-to-digital and the digital-to-analog converters. In a numerical simulation, $\delta$ can be set by the solver of the ordinary differential equations,~\footnote{The complete cycle of the diagram in Fig.~\ref{fig:diagram} is a multiple of $\delta$.} effectively setting the smallest time-scale and a natural time unit in the results section below.

A particular number $M$ of the most recent points read from the Environment (to be used in the consequent averaging of the mean field $X$) should be chosen according to the given characteristic time-scale of the collective mode. A set of points corresponding to multiple period of oscillations for slow control algorithms, or a single period for the faster ones, could both be logical choices in the Current State block to provide proper feedback.

\subsection{Action Block}
\label{sec:action}

Generally, action can be of any complicated waveform. Here, we restrict our consideration to the case of pulse train action, with the constant 
interval $\Delta$ between the pulses and an amplitude tuned for each pulse.
Taking into account that the pulse amplitude shall be limited by a 
value $A_{max}$ specific to a given application, we write our action as 
$${\cal C}(t) = \sum_n A_n{\cal P}(t_n)\,,$$
with $-A_{max}\leq A_n\leq A_{max}$ and $t_n=n\Delta$, $n=1,2,\ldots$. Here ${\cal P}(t_n)$ 
is a pulse of amplitude $A_n$ generated at the instant $t_n$.
Notice that $\Delta$ is a multiple of $\delta$.  

\begin{figure}
\centering
\includegraphics[width=0.75\columnwidth]{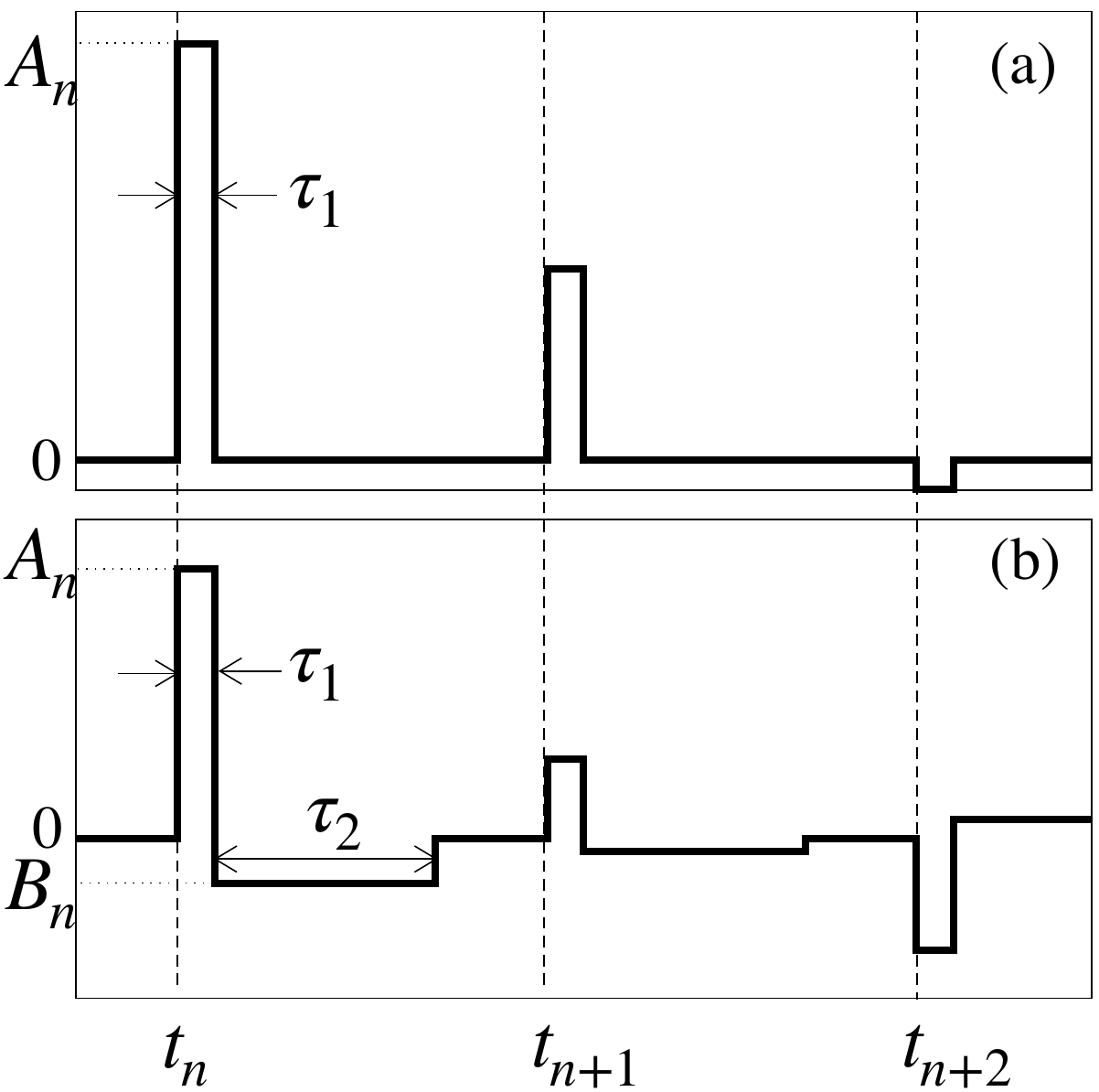}
\caption{Two types of pulses used for simulation.  The pulses appear at instants $t_n$ with 
the interval $\Delta$. The pulse amplitude $A_n$ is determined by the neural network for each pulse; notice that it can be both positive or negative. 
\textbf{(a)} Rectangular pulses have fixed width $\tau_1$. \textbf{(b)} Charged-balanced pulses 
are determined by two fixed parameters $\tau_{1,2}$; the second amplitude parameter 
$B_n$ is set to fulfil the condition $A_n\tau_1+B_n\tau_2=0$.
}
\label{fig:pulseform}
\end{figure}

We exploit pulses of two different forms. 
The first one is a simple rectangular pulse (Fig.~\ref{fig:pulseform}a)
\begin{align*}
 {\cal P}(t_n) &=
  \begin{cases}
   1        & \text{if } t_n\le t < t_n+\tau_1\,,  \\
   0        & \text{otherwise}\,.
  \end{cases}
\end{align*}
Next, because possible neuroscience applications require the pulses to be bipolar and charge-balanced, we consider the pulses given by 
\begin{align*}
 {\cal P}(t_n) &=
  \begin{cases}
   1        & \text{if } t_n\le t < t_n+\tau_1\,,  \\
   -\tau_1/\tau_2        &   \text{if } t_n+\tau_1\le t < t_n+\tau_1+\tau_2\,,\\
   0        & \text{otherwise}\,
  \end{cases}
\end{align*}
and illustrated in Fig.~\ref{fig:pulseform}b.~\footnote{
The currently available DBS devices deliver a pulsatile stimulation with a frequency of about 120 Hz. In particular, the pulse shape shown in Fig.~\ref{fig:pulseform}(b) is used.}

Naturally, smaller values of  $A_{max}$ are commonly sought after in biological applications such 
as the DBS in Parkinson's disease. 
Being physically just a current, the action is deliberately applied to the first equation in 
both~(\ref{eq:bvdp}) and (\ref{eq:HRn}), because it is that equation that describes the voltage across the neuron's membrane. The total ``energy'' per time supplied to the ensemble by $N_s$ pulses
\begin{equation}
A_\text{total}=(N_s\Delta)^{-1}\sum_{n=1}^{N_s} |A_n|\;,
\label{Atotal}
\end{equation}
is another measure that one aims to minimize in practice. 

Similar to other control problems, the immediate feedback from the Environment is desired. For that, we use a Runge--Kutta-based solver to evaluate Eqns.~(\ref{eq:bvdp}) or (\ref{eq:HRn}) numerically, provided the input to the Environment is equal to the action ${\cal C}(t)$. The Current State block extracts the current value of the mean field, $X(t_n)$, from the solver, imitating a measurement. 
To reflect a practical experimental realization, the PPO optimization will then rely only on the mean field array $X_{\text{state}}$ that contains a set of size $M$ of the most recent values of the mean field $X$, typically corresponding to a single period. 
In other words, it is assumed that one can measure variable reflecting collective oscillation of the ensemble in practice.

\subsection{Reward Block}
In this work we are interested in intuitive reward functions that can be easily interpreted physically. Therefore, for a given action with the pulse amplitude $A_n$ and a given array of observations 
$X_{\text{state}}$ at time $t_n$, we propose the following class of reward functions for synchrony suppression tasks: 
\begin{equation}\label{eq:reward-general}
R\big[t_n\big] =-\sum_{j=0}^2 \alpha_j \big(X(t_n)-\langle X_{\text{state}}\rangle_n\big)^j - \beta|A_n|,
\end{equation}
where the sum term rewards convergence of the system to an average of the mean field over previous 
$M$ values, $\langle X_{\text{state}}\rangle_n=M^{-1}\sum_{l=1}^M X(t_{n-l+1})$, 
and the second term favors smaller values of the stimulus amplitude $A_n$. 
Coefficients $\alpha_j$ and $\beta$ are the weight factors that, depending on an application, can introduce a bias towards a desired outcome (\textit{e.g.}, a more accurate convergence to a particular value of the mean field $X$ \emph{vs.} a smaller amplitude of the suppression pulse). Coefficients $\alpha_{1,2}$ define trajectory(rate) of desired suppression to the special point. The case of $\alpha_{1,2}$ = 0, $\alpha_0 > 0$ rewards for gradual decrease of stimuli to the baseline $A_n$=0.
Hence, the first term is a second order polynomial expanded around the target mean field value and the second term makes the action minimal while still suppressing the oscillations. We have the freedom to choose any reward function, but we chose the one that explicitly minimizes the stimuli amplitude.

Naturally, the Bonhoeffer-van der Pol ensemble of neurons~(\ref{eq:bvdp}) should be able to converge to the fixed point $\{X_0,Y_0\}$ if the system is rewarded simply as:
\begin{equation}\label{eq:reward-bvdp}
R_{\text{bvdp}}\big[t_n\big] = -\big(X(t_n)-\langle X_{\text{state}}\rangle_n\big)^2 - 2|A_n|,
\end{equation}
\noindent without the need to specify the values of $\{X_0,Y_0\}$ explicitly. We find empirically that this convergence is easily achievable in the regular model and the trade-off factor, responsible for minimization of the suppression pulse magnitude, could be increased to $\beta=2$ without any penalty.
On the other hand, the bursting oscillations of the Hindmarsh-Rose neurons impose an increased role on the second term in Eq.~\ref{eq:reward-general} (or a smaller parameter $\beta$ in the reward):
\begin{equation}\label{eq:reward-bursting}
R_{\text{burst}}\big[t_n\big]  = -2\big(X(t_n)-\langle X_{\text{state}}\rangle_n\big)^2 - 0.01 |A_n|.
\end{equation}

How to find the reward function automatically is the subject of active research in the machine learning community today~\cite{RL-rewardshaping}. For the scope of this manuscript, however, we restrict ourselves to the intuitive approaches of arriving at the coefficients $\alpha$ and $\beta$ in Eqns.~(\ref{eq:reward-bvdp}) and (\ref{eq:reward-bursting}). The problem of finding the reward automatically in the task of synchrony suppression and the corresponding strict optimization problem are yet to be solved.

\subsection{Proximal Policy Optimization Block}
We can now formally define our policy as a differentiable function (for brevity below we omit
index $n$ for the discrete time $t_n$):
\begin{equation}
\pi = \pi_\theta(X_{\text{state}},A)=\mathbb{P}_\theta\big\{X(t)=X_{\text{state}},\hspace{1pt} A(t)=A_n\big\},
\end{equation}
meaning that the policy $\pi$ is the probability of taking action $A_n$ when at state $X_{\text{state}}$ and the parameters approximating the action-state relationship are described by a vector $\theta$. Typically, $\theta$ represents weights of an artificial neural network that translates the values of an observed signal to an output action~\cite{RL-Sutton1998,RL-PPO-schulman2017}. The weights $\theta$ are initialized randomly and are then updated interatively as the network learns the proper translation. 

In policy-based Reinforcement Learning we want to maximize our total Reward Function, 
 \begin{equation}\mathcal{R}_\pi(\theta) =\mathbb{E}_{\pi}\Big[\sum^{T}_{t=0}\gamma^{t}\hspace{1pt}R\big[t\big]\Big],
 \label{eq:reward-Total}
 \end{equation}
which  shows  how  good  the  policy $\pi_{\theta}$ is  during  the  entire  synchrony  suppression  cycle of duration $T$ \footnote{$T$ is a macroscopic time-scale equal to the duration of stimuli application. For prospective wearable DBS systems, $T$ could be considered infinitely large.}. In Eq.~(\ref{eq:reward-Total}), $\mathbb{E}_{\pi}$ is a probabilistic expectation value, $\gamma$ is a discount factor that controls the trade-off between the long-term rewards and the immediate ones (typically equals to 0.99), and $R\big[t\big]$ is the reward at time $t$ described by any of the desired functions in Eqns.~(\ref{eq:reward-general})--(\ref{eq:reward-bursting}).

Following each measurement $X_{\text{state}}$ in the Current State block, the main idea of the PPO block is to be able to determine which action to take in order to maximize the reward given by Eq.~(\ref{eq:reward-Total}). The way to achieve this is to optimize the vector of parameters $\theta$.

The optimization follows Proximal Policy Optimization (PPO) algorithm that allows to choose optimal parameters to ascend in a direction of gradient in the policy, and towards maximal reward~\cite{RL-PolicyTheorem, RL-Sutton1998}. 
Taking derivative of Eq.~(\ref{eq:reward-Total}) and using policy gradient theorem~\cite{RL-PolicyTheorem},
one can factor out a Score Function term $\bigtriangledown_\theta \log \pi_\theta$ and an Advantage Function term $\mathbb{A}^\pi$: 
\begin{equation}
\bigtriangledown_\theta\mathcal{R}_\pi(\theta)=\mathbb{E}_{\pi \theta}\Big[\bigtriangledown_\theta \log \pi_\theta\big(X_{\text{state}},A_n\big)\hspace{2pt}\mathbb{A}^{\pi}\big(X_{\text{state}},A_n\big)\Big].
\label{eq:policygradient}
\end{equation}
It is these two factored terms in Eq.~(\ref{eq:policygradient}) that are evaluated by two artificial neural networks called ``Actor'' and ``Critic''. 
The latter, Critic, is a neural network that estimates Advantage function $\mathbb{A}^{\pi}\big(X_{\text{state}},A_n\big)$ at the current time $t$. In simple words, the Advantage function is nothing but a deviation of the current reward from the maximal prior reward recorded during the experiment up until the time moment $t$; and the Critic network builds a fit approximating its value.  
 This estimation is based on the reward function from Eq.~(\ref{eq:reward-general}), which will account both for the action optimization (minimize stimuli amplitude $A_n$) and for the state optimization (minimize $X(t)-X_{\text{state}}$) within the neural network. 
 The Actor is the other neural network that predicts the probability of actions for all states and updates the policy distribution according to the suggestion by the Critic~\cite{RL-Sutton1998,RL-PPO-schulman2017,RL-stable-baselines}. Weights of both artificial networks $\theta$, the policy $\pi_\theta$, and the Advantage function $\mathbb{A}^\pi$ are updated after each complete cycle of the diagram in Fig.~\ref{fig:diagram}. We selected PPO algorithm because it has proven to be both fast and robust, providing a simple framework for optimizing all hyper-parameters in the neural networks and for implementing parallelization of computation~\cite{RL-Sutton1998,RL-PPO-schulman2017}.

\section{Results}

\subsection{Experimental configuration}

All calculations were carried out on a CPU of a Desktop computer (Intel(R) Core(TM) i7-6700K CPU @ 4GHz, quad-core 64-bit x86 architecture, 8GiB DIMM DDR4, GeForce GTX 670). Within the Environment block, we used the standard Runge--Kutta algorithm to solve the differential equations numerically. For proper resolution, we chose the integration step $\delta = 0.2$ in the Bonhoeffer-van der Pol neuronal model and $\delta = 0.1$ in the bursting Hindmarsh-Rose neuronal model. Specific parameters of suppression pulses $\tau_{1,2}$ and $\Delta$ (and the duration of suppression $T$) are provided in the Figure captions in units of $\delta$. 

\begin{figure*}
\centering
\includegraphics[width=1.5\columnwidth]{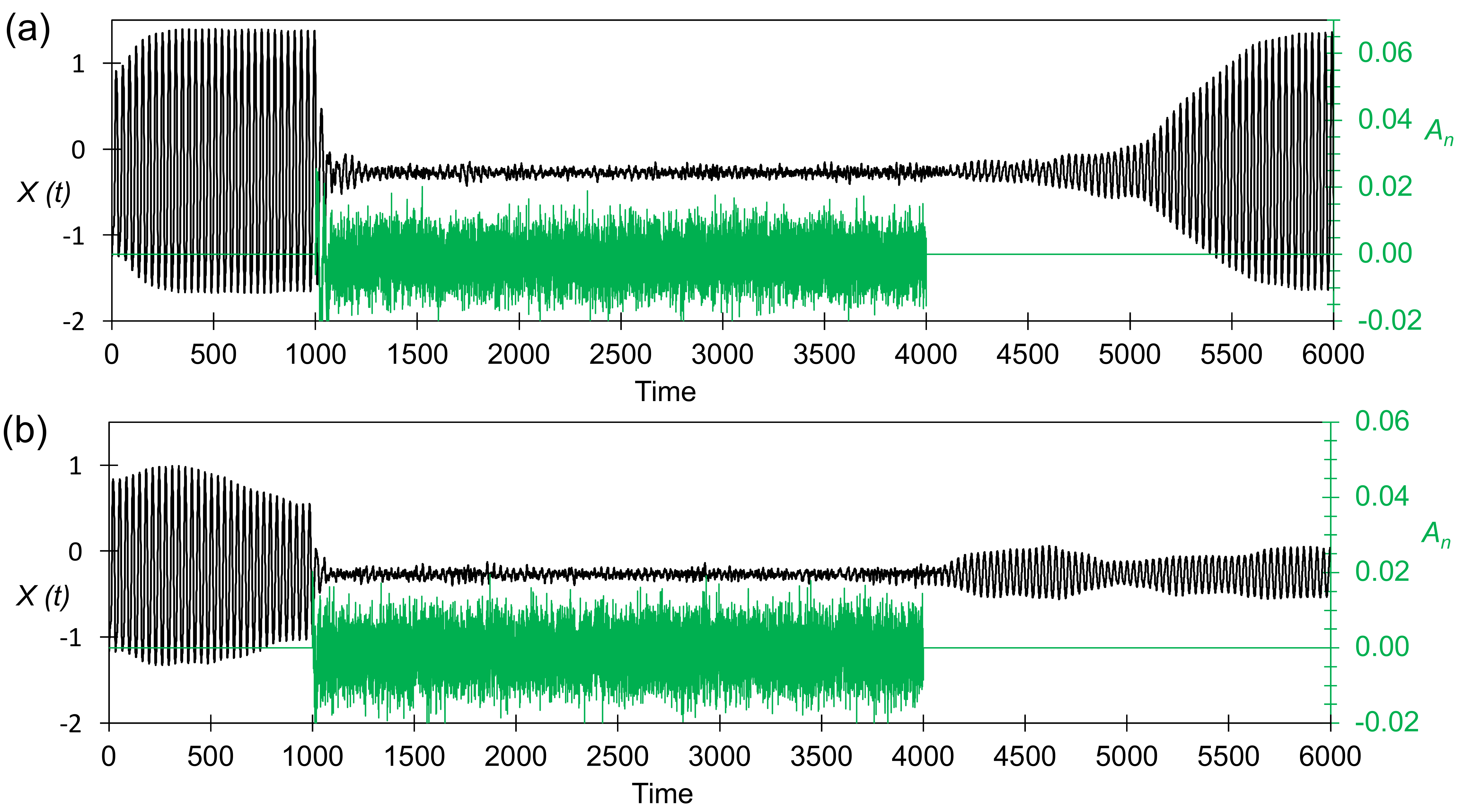}
\caption{Suppression of synchrony in Bonhoeffer-van der Pol neurons (\ref{eq:bvdp}) for $N=1000$, $M$ = 160 using rectangular pulses ($\tau_1$ = 1, $\Delta$ = 5). \textbf{(a)} Coupling strength $\e = 0.03$. \textbf{(b)} Coupling strength $\e = 0.02$, chaotic regime for the collective mode. Action pulses used for suppression (bottom green curves) are plotted against the right axis in the units of the mean field.
Notice that in this and in the following figures, the time intervals $\tau_{1,2}$ and $\Delta$ are given in the units of integration step $\delta$.
}
\label{fig:TS-chaos}
\end{figure*}

We took a reliable library of stable releases of the PPO/A2C algorithm~\cite{RL-stable-baselines} and used an approach called PPO-Clip\cite{RL-PPO-schulman2017} that allows not to leave the vicinity of the original policy $\pi_\theta$ when an update is calculated. Both the Critic and the Actor were parameterized with dense neural networks in the Tensorflow~\cite{tensorflow2015-whitepaper} framework. Each network contained 2 hidden layers with 64 artificial neurons and with the \emph{tanh} function as the activation between the hidden layers. We fine-tuned the PPO-Clip to run with the following parameters: 

\begin{center}
\begin{tabular}{ l l r l}
     $\bullet$ &\emph{gamma} = 0.99  & $\quad\bullet$ &\emph{n\_steps} = 128 \\
     $\bullet$ &\emph{ent\_coef} = 0.01 & $\quad\bullet$ &\emph{learning\_rate} = 0.00025 \\
     $\bullet$ &\emph{vf\_coef} = 0.5 &  $\quad\bullet$ &\emph{max\_grad\_norm}=0.5 \\
     $\bullet$ &\emph{lam} = 0.95 & $\quad\bullet$ &\emph{nminibatches} = 4 \\
     $\bullet$ &\emph{noptepochs} = 4  &  $\quad\bullet$ &\emph{cliprange} = 0.2 \\
\end{tabular}
\end{center}

During the stage of training the neural networks, the major optimization effort involved the step parameter \emph{n\_steps} and the entropy coefficient \emph{ent\_coef}. The former helps to find an optimal rate of updating the weights $\theta$ in the neural networks so that the solver has enough time to generate a substantially different output. The latter helps to reach the learning plateau faster by stochastically perturbing the current policy and effectively allowing it to reach its optima sooner.

The other parameters were kept at default values and are meant to provide a fast and robust convergence of the PPO algorithm. Despite quick convergence, the training was fixed to last 10 million steps to provide consistency in different experiments. The stimuli actions were restricted to take values in the range $-1\leq A_{max}\leq 1$. The values of errors reported below were calculated as a standard deviation in 16 repeated independent experiments.

\begin{figure*}
\centering
\includegraphics[width=1.55\columnwidth]{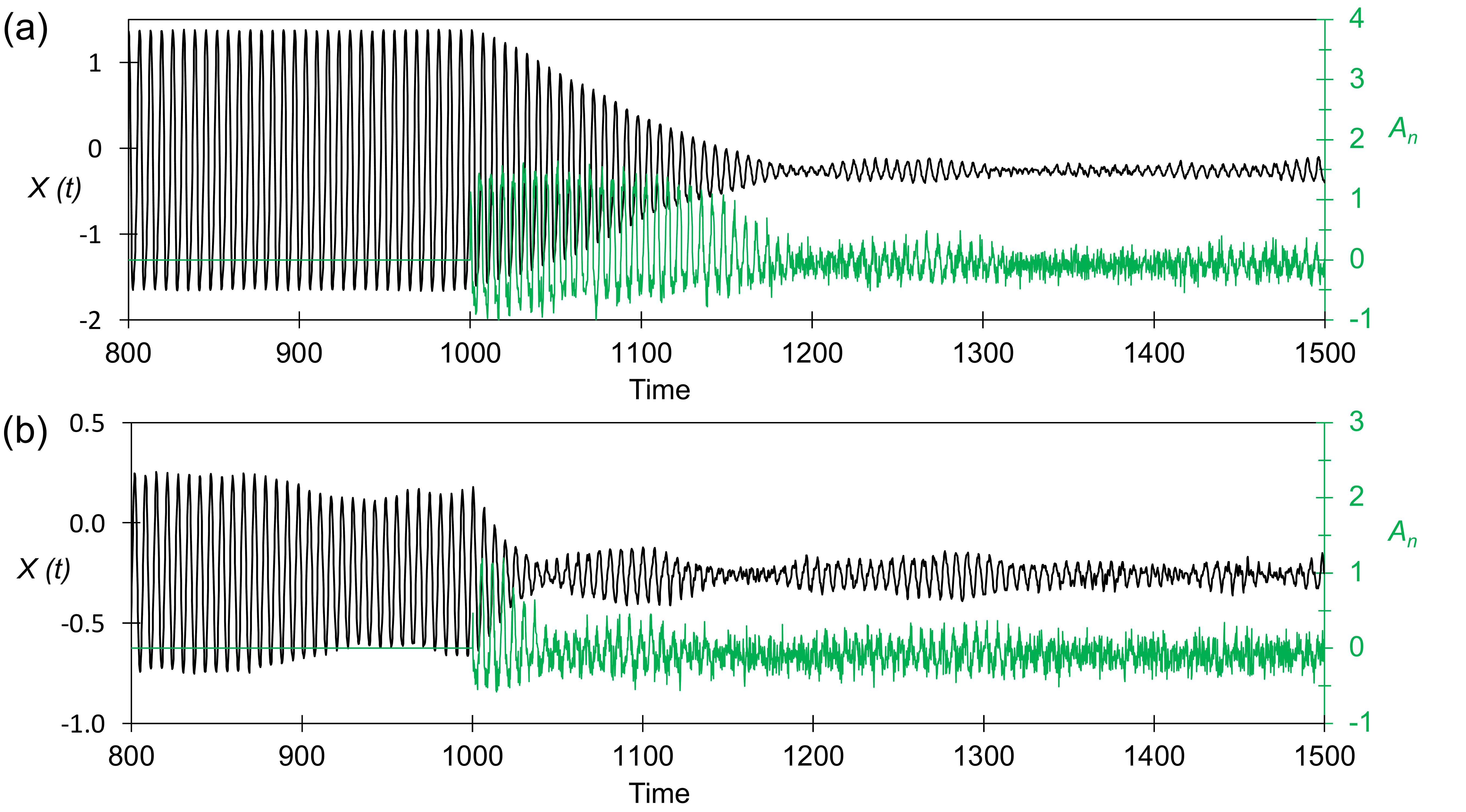}
\caption{Suppression of synchrony in Bonhoeffer-van der Pol neurons (\ref{eq:bvdp}) for $N=1000$, $M$ = 160 using charge-balanced pulses ($\tau_1$ = 1, $\tau_2 = 2$, $\Delta$ = 2), zoomed to the start of suppression. \textbf{(a)} Coupling strength $\e = 0.03$. \textbf{(b)} Coupling strength $\e = 0.02$, chaotic regime for the collective mode. Action pulses used for suppression (bottom green curves) are plotted against the right axis in the units of the mean field.
}
\label{fig:TS-CB}
\end{figure*}

\subsection{Synchrony suppression}
We first test the model on the ensemble of $N=1000$ self-sustained   Bonhoeffer-van der Pol neurons oscillating around a non-zero equilibrium point and globally coupled with $\e = 0.03$ (Fig.~\ref{fig:TS-chaos}(a)). 
We selected $M=160$ last points for calculating the reward functions because it corresponds to a single period of $X(t)$, but we also studied the dependence on M experimentally (see Fig.~\ref{fig:sup-coef} below).
At $t=1000$ we initiate synchrony suppression by sending action pulses according to the pre-trained Actor-Critic duet of neural networks. We observe that the reward function described by Eq.~(\ref{eq:reward-bvdp}) rewards both the convergence to the natural equilibrium point, yielding a non-zero average {$X_0\approx -0.2676 \pm 0.0010$}, 
and the smaller values of the action amplitudes, yielding almost two orders of magnitude reduction in the action amplitude (right axis in Fig.~\ref{fig:TS-chaos}(a)). Immediately after $t=1000$ the action values are somewhat large; however, they become of the order of $\sim 0.02$ in about 300 time steps.
We also observe gradual relaxation of the ensemble to the original state as the stimulation is switched off at $t=4000$.

\begin{figure*}
\includegraphics[width=1.6\columnwidth]{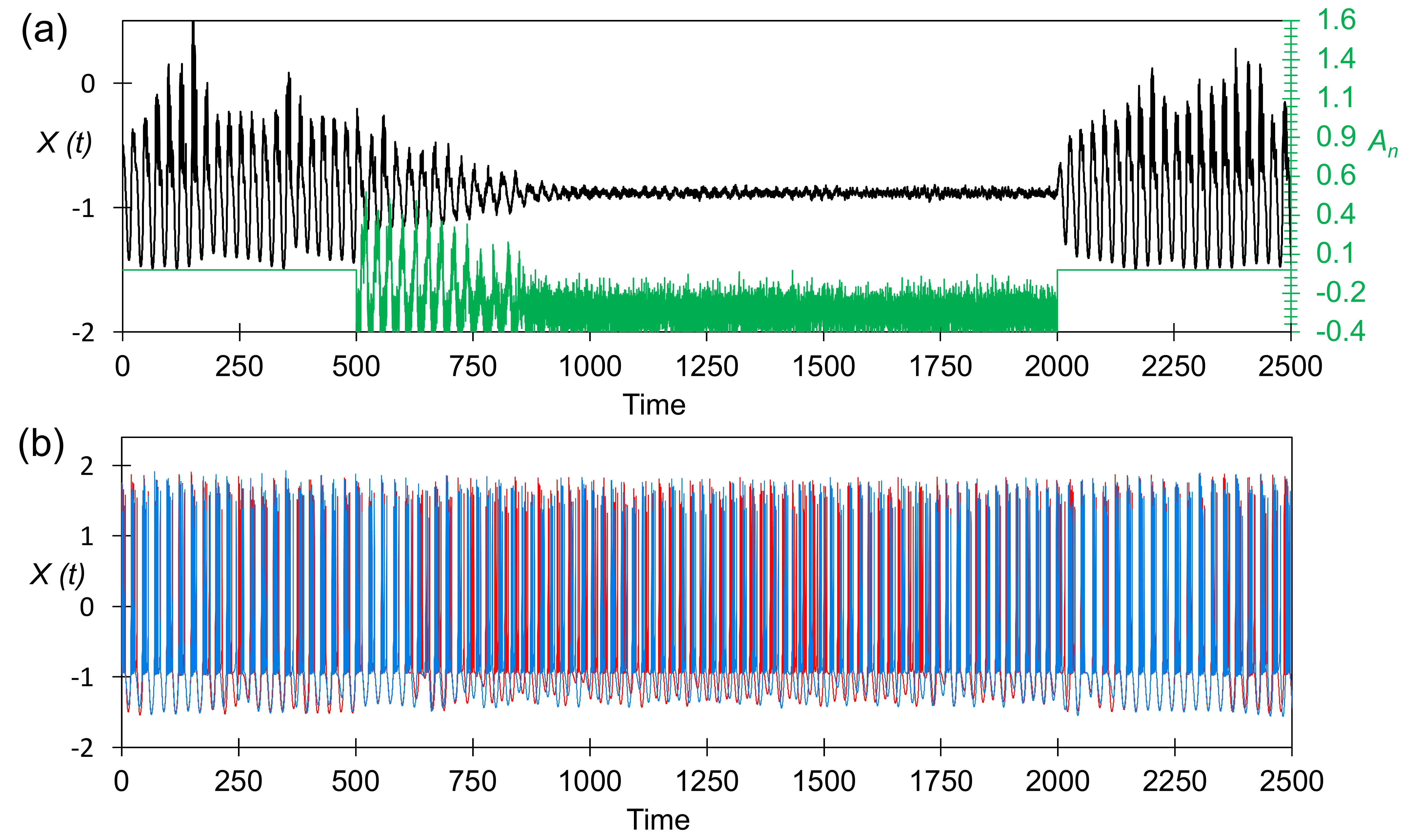}
\caption{Suppression of synchrony in bursting Hindmarsh-Rose neurons (\ref{eq:HRn}) for $\e=0.2$, $N=5000$, $M$ = 160 with rectangular pulses ($\tau_1$ = 1, $\Delta$ = 5). \textbf{(a)} Mean field (top black curve) and action pulses used for suppression (bottom green curve, plotted against the right axis in the same units as the mean field). \textbf{(b)} Dynamics of two randomly chosen neurons illustrates that suppression of the collective mode is accompanied by desynchronization of individual units.}
\label{fig:Bursting-TS}
\end{figure*}
\begin{figure*}
\includegraphics[width=1.55\columnwidth]{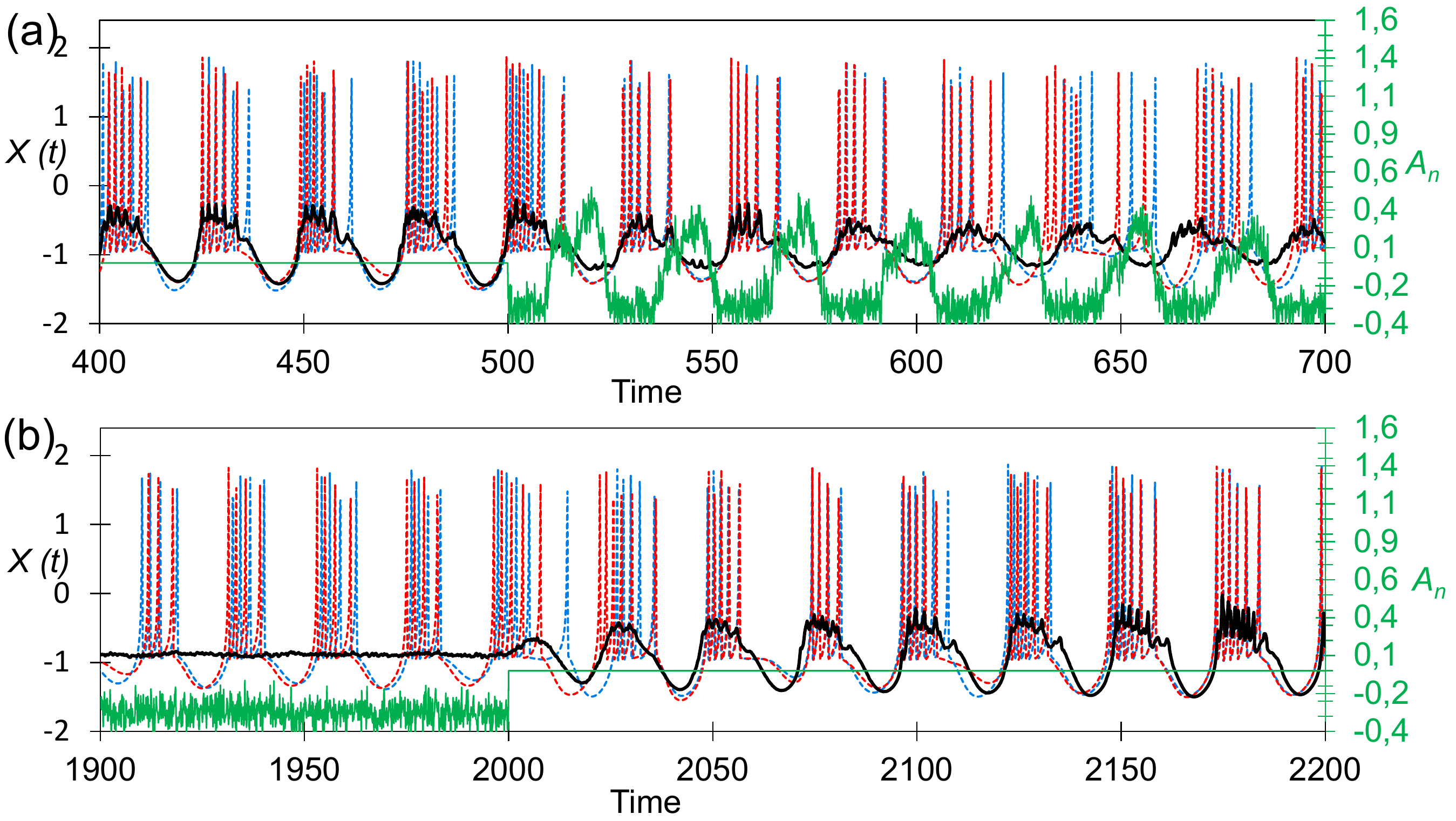}
\caption{Zoomed regions from FIG.~\ref{fig:Bursting-TS}. \textbf{(a)} Initiation of suppression at $t = 500$. \textbf{(b)} Relaxation after the suppression is switched off at $t=2000$. Dotted curves are two different neurons. Thick curve is the mean field. Thin curves are the action pulses used for suppression, all of which are plotted against the right axis in the same units as the mean field.}
\label{fig:Bursting-TS-zoomed}
\end{figure*}

Reinforcement learning can also suppress synchronization in the Bonhoeffer-van der Pol ensemble 
when the collective mode is chaotic ($\e = 0.02$). The results are shown in Fig.~\ref{fig:TS-chaos}(b). Although the oscillatory dynamics is now irregular, the Actor-Critic algorithm (rewarded with the same Eq.~(\ref{eq:reward-bvdp})) performs here similarly to the non-chaotic regimes, with {$X_0\approx -0.2596 \pm 0.0001$}, the same order of magnitude of the required amplitudes ($\sim 0.02$).
Naturally, relaxation to the original state of the ensemble in the chaotic regime, when the stimuli are switched off, occurs slower than in the regular regime with stronger coupling.

Charge-balanced pulses also prove functional to suppress synchrony both in the regular (Fig.~\ref{fig:TS-CB}(a)) and in the chaotic (Fig.~\ref{fig:TS-CB}(b)) regimes. The ensemble reaches the special point {$X_0\approx -0.2673 \pm 0.0029$} in the former and {$X_0\approx -0.2615 \pm 0.0010$} in the latter case.

Next, we consider the case of Hindmarsh-Rose neurons (\ref{eq:HRn}) for $\e=0.2$ and $N=5000$. The results of synchrony suppression, given a reward function Eq.~(\ref{eq:reward-bursting}), are shown in Figs.~\ref{fig:Bursting-TS} and ~\ref{fig:Bursting-TS-zoomed}. The bursting nature of the oscillators is seen in Fig.~\ref{fig:Bursting-TS}(b) 
and its zoomed version Fig.~\ref{fig:Bursting-TS-zoomed}.
After a series of action pulses is applied at $t=500$, 
the ensemble undergoes through a transient regime of about 400 time units and eventually desynchronizes.
This becomes obvious when one observes two randomly chosen individual neurons from the ensemble (see Fig.~\ref{fig:Bursting-TS}(b) and Fig.~\ref{fig:Bursting-TS-zoomed}(a) for a zoomed region around $t=500$). 

As the Actor-Critic networks continue to adapt to the current state, the synchrony of oscillations vanishes around $t=600$ and the mean field converges to the special point {$X_0=-0.8754 \pm 0.0156$}. 
Ultimately, when the suppression stimuli are switched off at $t=2000$, the ensemble enters the transient regime again, with initial enhancement and consequent weakening of synchrony as the mean field $X(t)$ gradually relaxes back to the original pattern, see Fig.~\ref{fig:Bursting-TS-zoomed}(b). In addition to rectangular-shaped stimuli, we observe a similar order of magnitude suppression in the Hindmarsh-Rose ensemble with charge-balanced pulses ($\tau_1$ = 1, $\tau_2$ = 5, $\Delta$ = 8) yielding $X_0=-0.8756 \pm 0.0003$.

\subsection{Quantitative analysis}
\label{sec:discussion}

\begin{figure}
\includegraphics[width=1\columnwidth]{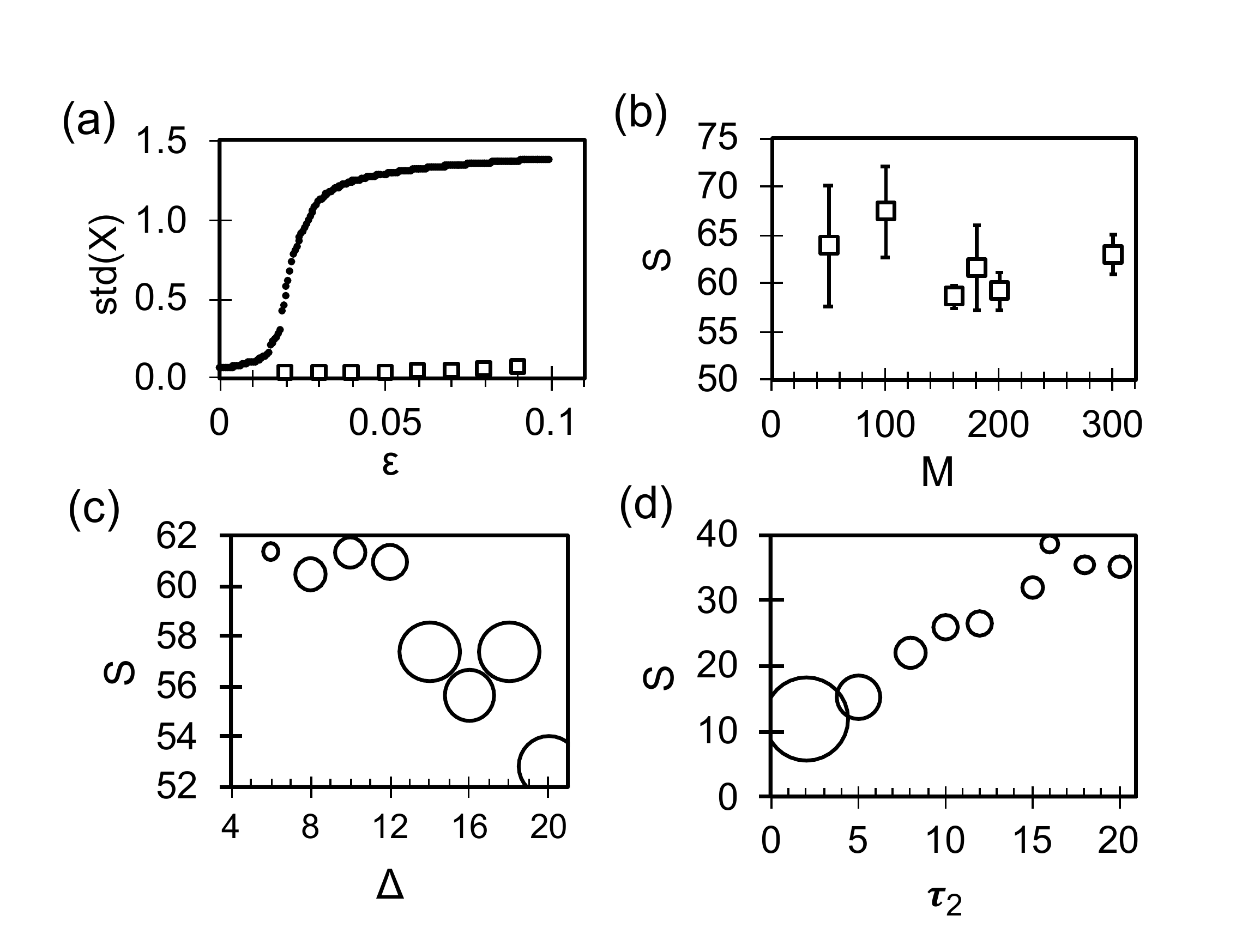}
\caption{Quantitative analysis of suppression in Bonhoeffer-van der Pol ensemble via RL. \textbf{(a)} Standard deviation (std) of the mean field $X$ \emph{vs} coupling strength $\e$. Dots show dependence before suppression and boxes show std values after transient relaxation, when a steady suppressed state is achieved by virtue of \textit{rectangular} pulses. \textbf{(b)} Suppression coefficient as a function of length of observation $M$. \textbf{(c)} Suppression coefficient as a function of distance  $\Delta$ between stimuli of rectangular shape.
\textbf{(d)} Suppression coefficient as a function of pulse width $\tau_2$ in case of \textit{charge-balanced} pulses. 
Error bars are standard deviations calculated on 16 experiments.
Bubble sizes are proportional to the total supplied energy $A_\text{total}$ (see Eq.~(\ref{Atotal}) for definition).
}
\label{fig:sup-coef}
\end{figure}

In order to get additional insight into the efficacy of the RL-based suppression, we studied its dependence on various parameters of the system and of the stimulation.
The major factor that determines the amplitude of the collective oscillation is the coupling strength $\e$, which we thoroughly varied. The result for the Bonhoeffer -- van der Pol ensemble, Eq.~(\ref{eq:bvdp}), is shown in Fig.~\ref{fig:sup-coef}. 

For the unperturbed system, the dependence of the standard deviation of the collective mode, $\text{std}(X_{\text{uncontrolled}})$, on the coupling strength $\e$ follows a threshold-like curve, typical for the Kuramoto transition~\cite{Kuramoto-84}, see dots in Fig.~\ref{fig:sup-coef}(a), while the $\text{std}(X_{\text{controlled}})$ for the suppressed field is shown in the same plot by boxes.
The latter quantity was computed when the Actor-Critic setup reached the best possible level of synchrony suppression. As can be seen in Fig.~\ref{fig:TS-chaos}, 
this final steady stage of the control is achieved soon after the stimuli application is switched on, at about $t_\text{steady}=1200$, and is preserved until the control is switched off at 
$t_\text{off}=4000$. (The corresponding values for the Hindmarsh-Rose model are $t_\text{steady}\approx 700$ and $t_\text{off}=2000$, see Fig.~\ref{fig:Bursting-TS-zoomed}). 

In the suppressed steady state the mean field continues to jitter due to the final size fluctuations. 
The amplitude of the stimuli also fluctuates but the pulse sequence now has a uniform variance and a diminished range of amplitudes required to keep the control active. 
The mean field fluctuations are known to depend on the ensemble size as~\cite{Pikovsky-Ruffo-99} $1/\sqrt{N}$.
Figure~\ref{fig:sup-coef}(a) suggests that the suppressed collective mode approximately reaches the finite size limit
~\footnote{The level of the finite-size fluctuations can be estimated as $\text{std}(X)$ for the sub-threshold values of the coupling $\e$.}.
We speculate that the additional source of fluctuations -- originating from the probabilistic uncertainty inherent to the Artificial Neural Networks that approximate the Score and the Advantage functions -- effectively trains the control algorithm to operate with noisy inputs, thus making it not only very efficient but also suitable for the experimental approbation in the future. 

The extent of suppression of the mean field prior to the stimuli application, $X_{\text{uncontrolled}}$, given the mean-field values after, $X_{\text{controlled}}$, can be quantified by the following suppression coefficient 
$$S=\text{std}\big[X_{\text{uncontrolled}}\big]\Big/\text{std}\big[X_{\text{controlled}}\big]\,.$$ 
Because the fluctuations of the suppressed field practically do not depend on $\e$, but the amplitude of the collective 
mode of the unperturbed filed grows with $\e$, see Fig.~\ref{fig:sup-coef}(a), the suppression coefficient is maximal for strongly synchronized systems and achieves $S = 65.815 \pm 2.472$ for the Bonhoeffer--van der Pol ensemble. In the Hindmarsh-Rose model, the system is suppressed to $S=20.509 \pm 0.710$  by rectangular pulses ($\tau_1=1, \Delta = 5$) and to $S=24.821 \pm 0.900$ by charge-balanced pulses ($\tau_1$ = 1, $\tau_2$ = 5, $\Delta$ = 8).

Figure~\ref{fig:sup-coef}bdemonstrates effect of the parameter $M$ which is the number of the stored recent values of the mean field. We see that suppression coefficient $S$ is not too sensitive with respect to variation 
of $M$ around the value corresponding to the period of the field to be controlled. Notably, the spread of error bars on this dependence reduces around the single period value  $M$ = 160, suggesting a stable local minimum for optimizing the other parameters.

We have also studied the effect of pulse intervals $\Delta$ and $\tau_2$ on the suppression efficiency (Fig.~\ref{fig:sup-coef}c,d). The rationale behind this test is to look for the optimal frequency of action pulses in order to minimize the energy of the perturbation but to still suppress the synchrony.
 The resulting fall-off in the suppression efficiency in Figs.~\ref{fig:sup-coef}(c) and (d) can be deemed as a classic example of trade-off either when a limited stimuli energy $A_\text{total}$ must be used or when an incomplete suppression is desired.

\section{Discussion}
We demonstrated a successful adaptation of Reinforcement Learning to the synchrony suppression task in a globally coupled oscillatory network. Having considered limit-cycle  Bonhoeffer-van der Pol oscillators and bursting Hindmarsh-Rose neurons as the test models, the method proved functional both for regular and for chaotic collective oscillations, without having the knowledge about the ensemble model.
The suppression coefficient is of the same order of magnitude as the known feedback-based techniques~\cite{Popovych_et_al-17}, with the major differences occurring near the finite size fluctuations limit.
We foresee further enhancement of the performance with the development of the proposed RL-approach, along the following lines.

One option is to perform suppression via RL with architectures entailing deep artificial neural networks~\cite{RL-Deep-arulkumaran2017, Yeo-Melnyk-19} (instead of the two-layer ones considered herein). These approaches are known to be efficient when the networks of interacting neurons become especially large, as well as when a signal of interest is emitted at the same level as noise. As such, the deep control methods could be further enhanced by the signal-noise coupling techniques~\cite{dylov2011nonlinear, *dylov2011instability}, where the signal could be enhanced at expense of the noise thanks to the nonlinear nature of the environment~\cite{dylov2010natphotSR}. 

Another option for improvement of the reported results entails introduction of a \emph{secondary} Actor-Critic model~\cite{RL-Multiagent-busoniu2006multi}. 
In our preliminary experiment, we have trained such auxiliary model during the transient patterns such as those occurring immediately after $t=t_\text{on}$ in the suppressed regime.
We observe that when this secondary model overtakes the control, it further reduces the amplitude of the mean field $X$ and desynchronizes the ensemble beyond the performance of a single model (preliminary results show suppression improvement by 12\% compared to the single model). 
This works because the response of the globally coupled ensembles is, generally speaking, nonlinear with respect to the stimuli amplitude. 
Indeed, this response is determined by the corresponding phase response curve that does not depend on the stimuli amplitude only in the limit of an infinitely small action. 

Long-term, one could envision a library of such Artificial Neural Networks pre-trained at different amplitude levels and at different values of pulse parameters ($A_{max}$, $\tau_{1,2}$, and $\Delta$) -- all to be embedded into the software controlling a DBS device. This promises a personalized approach to the patients with different signalling patterns and at different progression stages of the disease, regardless of its aetiology. The online learning could then utilize such a library and actively update the pre-trained models using the feedback signal measured by the DBS device.

We emphasise that the cause-effect relationship between the synchrony and the pathology is still an unproved hypothesis in neurobiology and computational neuroscience.
Control approaches that do not rely on this hypothesis are therefore naturally of interest, with the agnostic algorithms such as the RL method presented here being the most promising solution~\footnote{Notice that feedback-based techniques reported in [\cite{Popovych_et_al-17}] also do not use any information about the model of the system but explicitly assume that pathological activity emerges due to synchrony. 
}.  Machine learning methods could be proposed for optimization of the stimulation parameters regardless of the aetiology of the disease. The way to demonstrate this before proceeding to the clinical experimentation is to go beyond our simplistic model 
and to consider complex brain models such as the virtual brain~\cite{TheVirtualBrain},
also with account of synaptic plasticity in order to study long-lasting effects of desynchronization \cite{Tass-Majtanik-06}, or a live animal brain model~\cite{ParkinsonAnimalModel}.

Important advantage of the RL-based suppression method is that it is data-driven and universal. It could be readily implemented in an experimental setting, if one takes the measuring/stimulating equipment characteristics and limitations into account. 
For example, currently available DBS devices are not capable of adjusting pulse amplitude sufficiently fast for the feedback-based technique, presented in Refs.~\cite{Popovych_et_al-17},~\footnote{Notice that the previously developed feedback-based suppression techniques were analyzed theoretically only for the unrealistic continuous-time stimulation and their modification to the pulsatile stimulation was an \emph{ad hoc}, relying on the assumption of a smooth continuous envelope.}. The suppression workflow proposed in diagram of Fig.~\ref{fig:diagram}, however, is universal and can be exploited for a predefined stimulation pattern. We find Reinforced Learning to be an ideal candidate for the clinical approbation as a ``smart'' control algorithm to be embedded into the deep brain stimulation devices.

The code used for simulations is available upon request from the authors.

\begin{acknowledgments}
DD acknowledges the support of IoT Center of Excellence of the National Technology Initiative of Russia.
\end{acknowledgments}


\begin{thebibliography}{53}%
\makeatletter
\providecommand \@ifxundefined [1]{%
 \@ifx{#1\undefined}
}%
\providecommand \@ifnum [1]{%
 \ifnum #1\expandafter \@firstoftwo
 \else \expandafter \@secondoftwo
 \fi
}%
\providecommand \@ifx [1]{%
 \ifx #1\expandafter \@firstoftwo
 \else \expandafter \@secondoftwo
 \fi
}%
\providecommand \natexlab [1]{#1}%
\providecommand \enquote  [1]{``#1''}%
\providecommand \bibnamefont  [1]{#1}%
\providecommand \bibfnamefont [1]{#1}%
\providecommand \citenamefont [1]{#1}%
\providecommand \href@noop [0]{\@secondoftwo}%
\providecommand \href [0]{\begingroup \@sanitize@url \@href}%
\providecommand \@href[1]{\@@startlink{#1}\@@href}%
\providecommand \@@href[1]{\endgroup#1\@@endlink}%
\providecommand \@sanitize@url [0]{\catcode `\\12\catcode `\$12\catcode
  `\&12\catcode `\#12\catcode `\^12\catcode `\_12\catcode `\%12\relax}%
\providecommand \@@startlink[1]{}%
\providecommand \@@endlink[0]{}%
\providecommand \url  [0]{\begingroup\@sanitize@url \@url }%
\providecommand \@url [1]{\endgroup\@href {#1}{\urlprefix }}%
\providecommand \urlprefix  [0]{URL }%
\providecommand \Eprint [0]{\href }%
\providecommand \doibase [0]{http://dx.doi.org/}%
\providecommand \selectlanguage [0]{\@gobble}%
\providecommand \bibinfo  [0]{\@secondoftwo}%
\providecommand \bibfield  [0]{\@secondoftwo}%
\providecommand \translation [1]{[#1]}%
\providecommand \BibitemOpen [0]{}%
\providecommand \bibitemStop [0]{}%
\providecommand \bibitemNoStop [0]{.\EOS\space}%
\providecommand \EOS [0]{\spacefactor3000\relax}%
\providecommand \BibitemShut  [1]{\csname bibitem#1\endcsname}%
\let\auto@bib@innerbib\@empty
\bibitem [{\citenamefont {Benabid}\ \emph {et~al.}(1991)\citenamefont
  {Benabid}, \citenamefont {Pollak}, \citenamefont {Gervason}, \citenamefont
  {Hoffmann}, \citenamefont {Gao}, \citenamefont {Hommel}, \citenamefont
  {Perret},\ and\ \citenamefont {De~Rougemont}}]{Benabid_et_al-91}%
  \BibitemOpen
  \bibfield  {author} {\bibinfo {author} {\bibfnamefont {A.}~\bibnamefont
  {Benabid}}, \bibinfo {author} {\bibfnamefont {P.}~\bibnamefont {Pollak}},
  \bibinfo {author} {\bibfnamefont {C.}~\bibnamefont {Gervason}}, \bibinfo
  {author} {\bibfnamefont {D.}~\bibnamefont {Hoffmann}}, \bibinfo {author}
  {\bibfnamefont {D.}~\bibnamefont {Gao}}, \bibinfo {author} {\bibfnamefont
  {M.}~\bibnamefont {Hommel}}, \bibinfo {author} {\bibfnamefont
  {J.}~\bibnamefont {Perret}}, \ and\ \bibinfo {author} {\bibfnamefont
  {J.}~\bibnamefont {De~Rougemont}},\ }\bibfield  {title} {\enquote {\bibinfo
  {title} {Long-term suppression of tremor by chronic stimulation of the
  ventral intermediate thalamic nucleus},}\ }\href@noop {} {\bibfield
  {journal} {\bibinfo  {journal} {Lancet}\ }\textbf {\bibinfo {volume} {337}},\
  \bibinfo {pages} {403--406} (\bibinfo {year} {1991})}\BibitemShut {NoStop}%
\bibitem [{\citenamefont {Benabid}\ \emph {et~al.}(2009)\citenamefont
  {Benabid}, \citenamefont {Chabardes}, \citenamefont {Mitrofanis},\ and\
  \citenamefont {Pollak}}]{Benabid_et_al-09}%
  \BibitemOpen
  \bibfield  {author} {\bibinfo {author} {\bibfnamefont {A.}~\bibnamefont
  {Benabid}}, \bibinfo {author} {\bibfnamefont {S.}~\bibnamefont {Chabardes}},
  \bibinfo {author} {\bibfnamefont {J.}~\bibnamefont {Mitrofanis}}, \ and\
  \bibinfo {author} {\bibfnamefont {P.}~\bibnamefont {Pollak}},\ }\bibfield
  {title} {\enquote {\bibinfo {title} {Deep brain stimulation of the
  subthalamic nucleus for the treatment of {P}arkinson's disease},}\
  }\href@noop {} {\bibfield  {journal} {\bibinfo  {journal} {Lancet Neurol.}\
  }\textbf {\bibinfo {volume} {8}},\ \bibinfo {pages} {67--81} (\bibinfo {year}
  {2009})}\BibitemShut {NoStop}%
\bibitem [{\citenamefont {K\"uhn}\ and\ \citenamefont
  {Volkmann}(2017)}]{Kuehn-Volkmann-17}%
  \BibitemOpen
  \bibfield  {author} {\bibinfo {author} {\bibfnamefont {A.}~\bibnamefont
  {K\"uhn}}\ and\ \bibinfo {author} {\bibfnamefont {J.}~\bibnamefont
  {Volkmann}},\ }\bibfield  {title} {\enquote {\bibinfo {title} {Innovations in
  deep brain stimulation methodology},}\ }\href@noop {} {\bibfield  {journal}
  {\bibinfo  {journal} {Mov. Disorders.}\ }\textbf {\bibinfo {volume} {32}},\
  \bibinfo {pages} {11} (\bibinfo {year} {2017})}\BibitemShut {NoStop}%
\bibitem [{\citenamefont {Johnson}\ \emph {et~al.}(2008)\citenamefont
  {Johnson}, \citenamefont {Miocinovic}, \citenamefont {McIntyre},\ and\
  \citenamefont {Vitek}}]{Johnson2008}%
  \BibitemOpen
  \bibfield  {author} {\bibinfo {author} {\bibfnamefont {M.~D.}\ \bibnamefont
  {Johnson}}, \bibinfo {author} {\bibfnamefont {S.}~\bibnamefont {Miocinovic}},
  \bibinfo {author} {\bibfnamefont {C.~C.}\ \bibnamefont {McIntyre}}, \ and\
  \bibinfo {author} {\bibfnamefont {J.~L.}\ \bibnamefont {Vitek}},\ }\bibfield
  {title} {\enquote {\bibinfo {title} {Mechanisms and targets of deep brain
  stimulation in movement disorders},}\ }\href@noop {} {\bibfield  {journal}
  {\bibinfo  {journal} {Neurotherapeutics}\ }\textbf {\bibinfo {volume} {5}},\
  \bibinfo {pages} {294--308} (\bibinfo {year} {2008})}\BibitemShut {NoStop}%
\bibitem [{\citenamefont {Gradinaru}\ \emph {et~al.}(2009)\citenamefont
  {Gradinaru}, \citenamefont {Mogri}, \citenamefont {Thompson}, \citenamefont
  {Henderson},\ and\ \citenamefont {Deisseroth}}]{Gradinaru-09}%
  \BibitemOpen
  \bibfield  {author} {\bibinfo {author} {\bibfnamefont {V.}~\bibnamefont
  {Gradinaru}}, \bibinfo {author} {\bibfnamefont {M.}~\bibnamefont {Mogri}},
  \bibinfo {author} {\bibfnamefont {K.~R.}\ \bibnamefont {Thompson}}, \bibinfo
  {author} {\bibfnamefont {J.~M.}\ \bibnamefont {Henderson}}, \ and\ \bibinfo
  {author} {\bibfnamefont {K.}~\bibnamefont {Deisseroth}},\ }\bibfield  {title}
  {\enquote {\bibinfo {title} {Optical deconstruction of {P}arkinsonian neural
  circuitry},}\ }\href@noop {} {\bibfield  {journal} {\bibinfo  {journal}
  {Science}\ }\textbf {\bibinfo {volume} {324}},\ \bibinfo {pages} {354--359}
  (\bibinfo {year} {2009})}\BibitemShut {NoStop}%
\bibitem [{\citenamefont {Deniau}\ \emph {et~al.}(2010)\citenamefont {Deniau},
  \citenamefont {Degos}, \citenamefont {Bosch},\ and\ \citenamefont
  {Maurice}}]{Deniau_et_al-10}%
  \BibitemOpen
  \bibfield  {author} {\bibinfo {author} {\bibfnamefont {J.-M.}\ \bibnamefont
  {Deniau}}, \bibinfo {author} {\bibfnamefont {B.}~\bibnamefont {Degos}},
  \bibinfo {author} {\bibfnamefont {C.}~\bibnamefont {Bosch}}, \ and\ \bibinfo
  {author} {\bibfnamefont {N.}~\bibnamefont {Maurice}},\ }\bibfield  {title}
  {\enquote {\bibinfo {title} {Deep brain stimulation mechanisms: beyond the
  concept of local functional inhibition},}\ }\href@noop {} {\bibfield
  {journal} {\bibinfo  {journal} {European Journal of Neuroscience}\ }\textbf
  {\bibinfo {volume} {32}},\ \bibinfo {pages} {1080--1091} (\bibinfo {year}
  {2010})}\BibitemShut {NoStop}%
\bibitem [{\citenamefont {Tass}(1999)}]{Tass-99}%
  \BibitemOpen
  \bibfield  {author} {\bibinfo {author} {\bibfnamefont {P.~A.}\ \bibnamefont
  {Tass}},\ }\href@noop {} {\emph {\bibinfo {title} {Phase Resetting in
  Medicine and Biology. Stochastic Modelling and Data Analysis.}}}\ (\bibinfo
  {publisher} {Springer-Verlag},\ \bibinfo {address} {Berlin},\ \bibinfo {year}
  {1999})\BibitemShut {NoStop}%
\bibitem [{\citenamefont {Rosenblum}\ and\ \citenamefont
  {Pikovsky}(2004{\natexlab{a}})}]{Rosenblum-Pikovsky-04}%
  \BibitemOpen
  \bibfield  {author} {\bibinfo {author} {\bibfnamefont {M.~G.}\ \bibnamefont
  {Rosenblum}}\ and\ \bibinfo {author} {\bibfnamefont {A.~S.}\ \bibnamefont
  {Pikovsky}},\ }\bibfield  {title} {\enquote {\bibinfo {title} {Controlling
  synchrony in ensemble of globally coupled oscillators},}\ }\href@noop {}
  {\bibfield  {journal} {\bibinfo  {journal} {Phys. Rev. Lett.}\ }\textbf
  {\bibinfo {volume} {92}},\ \bibinfo {pages} {114102} (\bibinfo {year}
  {2004}{\natexlab{a}})}\BibitemShut {NoStop}%
\bibitem [{\citenamefont {Rosenblum}\ and\ \citenamefont
  {Pikovsky}(2004{\natexlab{b}})}]{Rosenblum-Pikovsky-04a}%
  \BibitemOpen
  \bibfield  {author} {\bibinfo {author} {\bibfnamefont {M.~G.}\ \bibnamefont
  {Rosenblum}}\ and\ \bibinfo {author} {\bibfnamefont {A.~S.}\ \bibnamefont
  {Pikovsky}},\ }\bibfield  {title} {\enquote {\bibinfo {title} {Delayed
  feedback control of collective synchrony: {A}n approach to suppression of
  pathological brain rhythms},}\ }\href@noop {} {\bibfield  {journal} {\bibinfo
   {journal} {Phys. Rev. E.}\ }\textbf {\bibinfo {volume} {70}},\ \bibinfo
  {pages} {041904} (\bibinfo {year} {2004}{\natexlab{b}})}\BibitemShut
  {NoStop}%
\bibitem [{\citenamefont {Popovych}, \citenamefont {Hauptmann},\ and\
  \citenamefont {Tass}(2005)}]{Popovych-Hauptmann-Tass-05}%
  \BibitemOpen
  \bibfield  {author} {\bibinfo {author} {\bibfnamefont {O.}~\bibnamefont
  {Popovych}}, \bibinfo {author} {\bibfnamefont {C.}~\bibnamefont {Hauptmann}},
  \ and\ \bibinfo {author} {\bibfnamefont {P.~A.}\ \bibnamefont {Tass}},\
  }\bibfield  {title} {\enquote {\bibinfo {title} {Effective desynchronization
  by nonlinear delayed feedback},}\ }\href@noop {} {\bibfield  {journal}
  {\bibinfo  {journal} {Phys. Rev. Lett.}\ }\textbf {\bibinfo {volume} {94}},\
  \bibinfo {pages} {164102} (\bibinfo {year} {2005})}\BibitemShut {NoStop}%
\bibitem [{\citenamefont {Tukhlina}\ \emph {et~al.}(2007)\citenamefont
  {Tukhlina}, \citenamefont {Rosenblum}, \citenamefont {Pikovsky},\ and\
  \citenamefont {Kurths}}]{Tukhlina-Rosenblum-Pikovsky-Kurths-07}%
  \BibitemOpen
  \bibfield  {author} {\bibinfo {author} {\bibfnamefont {N.}~\bibnamefont
  {Tukhlina}}, \bibinfo {author} {\bibfnamefont {M.}~\bibnamefont {Rosenblum}},
  \bibinfo {author} {\bibfnamefont {A.}~\bibnamefont {Pikovsky}}, \ and\
  \bibinfo {author} {\bibfnamefont {J.}~\bibnamefont {Kurths}},\ }\bibfield
  {title} {\enquote {\bibinfo {title} {Feedback suppression of neural synchrony
  by vanishing stimulation},}\ }\href@noop {} {\bibfield  {journal} {\bibinfo
  {journal} {Phys. Rev. E.}\ }\textbf {\bibinfo {volume} {75}},\ \bibinfo
  {pages} {011019} (\bibinfo {year} {2007})}\BibitemShut {NoStop}%
\bibitem [{\citenamefont {Montaseri}\ \emph {et~al.}(2013)\citenamefont
  {Montaseri}, \citenamefont {Javad~Yazdanpanah}, \citenamefont {Pikovsky},\
  and\ \citenamefont {Rosenblum}}]{Montaseri_et_al-13}%
  \BibitemOpen
  \bibfield  {author} {\bibinfo {author} {\bibfnamefont {G.}~\bibnamefont
  {Montaseri}}, \bibinfo {author} {\bibfnamefont {M.}~\bibnamefont
  {Javad~Yazdanpanah}}, \bibinfo {author} {\bibfnamefont {A.}~\bibnamefont
  {Pikovsky}}, \ and\ \bibinfo {author} {\bibfnamefont {M.}~\bibnamefont
  {Rosenblum}},\ }\bibfield  {title} {\enquote {\bibinfo {title} {Synchrony
  suppression in ensembles of coupled oscillators via adaptive vanishing
  feedback},}\ }\href@noop {} {\bibfield  {journal} {\bibinfo  {journal}
  {Chaos}\ }\textbf {\bibinfo {volume} {23}},\ \bibinfo {pages} {033122}
  (\bibinfo {year} {2013})}\BibitemShut {NoStop}%
\bibitem [{\citenamefont {Lin}\ \emph {et~al.}(2013)\citenamefont {Lin},
  \citenamefont {Pu}, \citenamefont {Guo},\ and\ \citenamefont
  {Kurths}}]{Lin_2013}%
  \BibitemOpen
  \bibfield  {author} {\bibinfo {author} {\bibfnamefont {W.}~\bibnamefont
  {Lin}}, \bibinfo {author} {\bibfnamefont {Y.}~\bibnamefont {Pu}}, \bibinfo
  {author} {\bibfnamefont {Y.}~\bibnamefont {Guo}}, \ and\ \bibinfo {author}
  {\bibfnamefont {J.}~\bibnamefont {Kurths}},\ }\bibfield  {title} {\enquote
  {\bibinfo {title} {Oscillation suppression and synchronization: Frequencies
  determine the role of control with time delays},}\ }\href@noop {} {\bibfield
  {journal} {\bibinfo  {journal} {{EPL} (Europhysics Letters)}\ }\textbf
  {\bibinfo {volume} {102}},\ \bibinfo {pages} {20003} (\bibinfo {year}
  {2013})}\BibitemShut {NoStop}%
\bibitem [{\citenamefont {Zhou}\ \emph {et~al.}(2017)\citenamefont {Zhou},
  \citenamefont {Ji}, \citenamefont {Zhou}, \citenamefont {Feng}, \citenamefont
  {Kurths},\ and\ \citenamefont {Lin}}]{Zhou_2017}%
  \BibitemOpen
  \bibfield  {author} {\bibinfo {author} {\bibfnamefont {S.}~\bibnamefont
  {Zhou}}, \bibinfo {author} {\bibfnamefont {P.}~\bibnamefont {Ji}}, \bibinfo
  {author} {\bibfnamefont {Q.}~\bibnamefont {Zhou}}, \bibinfo {author}
  {\bibfnamefont {J.}~\bibnamefont {Feng}}, \bibinfo {author} {\bibfnamefont
  {J.}~\bibnamefont {Kurths}}, \ and\ \bibinfo {author} {\bibfnamefont
  {W.}~\bibnamefont {Lin}},\ }\bibfield  {title} {\enquote {\bibinfo {title}
  {Adaptive elimination of synchronization in coupled oscillator},}\
  }\href@noop {} {\bibfield  {journal} {\bibinfo  {journal} {New Journal of
  Physics}\ }\textbf {\bibinfo {volume} {19}},\ \bibinfo {pages} {083004}
  (\bibinfo {year} {2017})}\BibitemShut {NoStop}%
\bibitem [{\citenamefont {Popovych}\ \emph {et~al.}(2017)\citenamefont
  {Popovych}, \citenamefont {Lysyansky}, \citenamefont {Rosenblum},
  \citenamefont {Pikovsky},\ and\ \citenamefont {Tass}}]{Popovych_et_al-17}%
  \BibitemOpen
  \bibfield  {author} {\bibinfo {author} {\bibfnamefont {O.}~\bibnamefont
  {Popovych}}, \bibinfo {author} {\bibfnamefont {B.}~\bibnamefont {Lysyansky}},
  \bibinfo {author} {\bibfnamefont {M.}~\bibnamefont {Rosenblum}}, \bibinfo
  {author} {\bibfnamefont {A.}~\bibnamefont {Pikovsky}}, \ and\ \bibinfo
  {author} {\bibfnamefont {P.}~\bibnamefont {Tass}},\ }\bibfield  {title}
  {\enquote {\bibinfo {title} {Pulsatile desynchronizing delayed feedback for
  closed-loop deep brain stimulation},}\ }\href@noop {} {\bibfield  {journal}
  {\bibinfo  {journal} {PLOS One}\ }\textbf {\bibinfo {volume} {12}},\ \bibinfo
  {pages} {e0173363} (\bibinfo {year} {2017})}\BibitemShut {NoStop}%
\bibitem [{\citenamefont {Tass}(2001)}]{Tass-01}%
  \BibitemOpen
  \bibfield  {author} {\bibinfo {author} {\bibfnamefont {P.~A.}\ \bibnamefont
  {Tass}},\ }\bibfield  {title} {\enquote {\bibinfo {title} {Effective
  desynchronization by means of double-pulse phase resetting},}\ }\href@noop {}
  {\bibfield  {journal} {\bibinfo  {journal} {Europhys Lett.}\ }\textbf
  {\bibinfo {volume} {53}},\ \bibinfo {pages} {15--21} (\bibinfo {year}
  {2001})}\BibitemShut {NoStop}%
\bibitem [{\citenamefont {Hauptmann}\ and\ \citenamefont
  {Tass}(2009)}]{Hauptmann-Tass-09}%
  \BibitemOpen
  \bibfield  {author} {\bibinfo {author} {\bibfnamefont {C.}~\bibnamefont
  {Hauptmann}}\ and\ \bibinfo {author} {\bibfnamefont {P.~A.}\ \bibnamefont
  {Tass}},\ }\bibfield  {title} {\enquote {\bibinfo {title} {Cumulative and
  after-effects of short and weak coordinated reset stimulation: a modeling
  study},}\ }\href@noop {} {\bibfield  {journal} {\bibinfo  {journal} {J Neural
  Eng.}\ }\textbf {\bibinfo {volume} {6}},\ \bibinfo {pages} {016004} (\bibinfo
  {year} {2009})}\BibitemShut {NoStop}%
\bibitem [{\citenamefont {V.Popovych}\ and\ \citenamefont
  {Tass}(2012)}]{Popovych-Tass-12}%
  \BibitemOpen
  \bibfield  {author} {\bibinfo {author} {\bibfnamefont {O.}~\bibnamefont
  {V.Popovych}}\ and\ \bibinfo {author} {\bibfnamefont {P.~A.}\ \bibnamefont
  {Tass}},\ }\bibfield  {title} {\enquote {\bibinfo {title} {Desynchronizing
  electrical and sensory coordinated reset neuromodulation.}}\ }\href@noop {}
  {\bibfield  {journal} {\bibinfo  {journal} {Front Hum Neurosci.}\ }\textbf
  {\bibinfo {volume} {6}},\ \bibinfo {pages} {58} (\bibinfo {year}
  {2012})}\BibitemShut {NoStop}%
\bibitem [{\citenamefont {Wilson}\ and\ \citenamefont
  {Moehlis}(2016)}]{Wilson-Moehlis-16}%
  \BibitemOpen
  \bibfield  {author} {\bibinfo {author} {\bibfnamefont {D.}~\bibnamefont
  {Wilson}}\ and\ \bibinfo {author} {\bibfnamefont {J.}~\bibnamefont
  {Moehlis}},\ }\bibfield  {title} {\enquote {\bibinfo {title} {Clustered
  desynchronization from high-frequency deep brain stimulation},}\ }\href@noop
  {} {\bibfield  {journal} {\bibinfo  {journal} {PLOS Computational Biology}\
  }\textbf {\bibinfo {volume} {11}},\ \bibinfo {pages} {1--26} (\bibinfo {year}
  {2016})}\BibitemShut {NoStop}%
\bibitem [{\citenamefont {Holt}\ \emph {et~al.}(2016)\citenamefont {Holt},
  \citenamefont {Wilson}, \citenamefont {Shinn}, \citenamefont {Moehlis},\ and\
  \citenamefont {Netoff}}]{Holt_et_al-16}%
  \BibitemOpen
  \bibfield  {author} {\bibinfo {author} {\bibfnamefont {A.}~\bibnamefont
  {Holt}}, \bibinfo {author} {\bibfnamefont {D.}~\bibnamefont {Wilson}},
  \bibinfo {author} {\bibfnamefont {M.}~\bibnamefont {Shinn}}, \bibinfo
  {author} {\bibfnamefont {J.}~\bibnamefont {Moehlis}}, \ and\ \bibinfo
  {author} {\bibfnamefont {T.}~\bibnamefont {Netoff}},\ }\bibfield  {title}
  {\enquote {\bibinfo {title} {Phasic burst stimulation: {A} closed-loop
  approach to tuning deep brain stimulation parameters for parkinson's
  disease},}\ }\href@noop {} {\bibfield  {journal} {\bibinfo  {journal} {PLoS
  Comput Biol.}\ }\textbf {\bibinfo {volume} {12}},\ \bibinfo {pages}
  {e1005011} (\bibinfo {year} {2016})}\BibitemShut {NoStop}%
\bibitem [{\citenamefont {Rosin}\ \emph {et~al.}(2011)\citenamefont {Rosin},
  \citenamefont {Slovik}, \citenamefont {Mitelman}, \citenamefont
  {Rivlin-Etzion}, \citenamefont {Haber}, \citenamefont {Israel}, \citenamefont
  {Vaadia},\ and\ \citenamefont {Bergman}}]{Rosin-11}%
  \BibitemOpen
  \bibfield  {author} {\bibinfo {author} {\bibfnamefont {B.}~\bibnamefont
  {Rosin}}, \bibinfo {author} {\bibfnamefont {M.}~\bibnamefont {Slovik}},
  \bibinfo {author} {\bibfnamefont {R.}~\bibnamefont {Mitelman}}, \bibinfo
  {author} {\bibfnamefont {M.}~\bibnamefont {Rivlin-Etzion}}, \bibinfo {author}
  {\bibfnamefont {S.~N.}\ \bibnamefont {Haber}}, \bibinfo {author}
  {\bibfnamefont {Z.}~\bibnamefont {Israel}}, \bibinfo {author} {\bibfnamefont
  {E.}~\bibnamefont {Vaadia}}, \ and\ \bibinfo {author} {\bibfnamefont
  {H.}~\bibnamefont {Bergman}},\ }\bibfield  {title} {\enquote {\bibinfo
  {title} {Closed-loop deep brain stimulation is superior in ameliorating
  parkinsonism},}\ }\href@noop {} {\bibfield  {journal} {\bibinfo  {journal}
  {Neuron}\ }\textbf {\bibinfo {volume} {72}},\ \bibinfo {pages} {370 -- 384}
  (\bibinfo {year} {2011})}\BibitemShut {NoStop}%
\bibitem [{\citenamefont {Little}\ \emph {et~al.}(2013)\citenamefont {Little},
  \citenamefont {Pogosyan}, \citenamefont {Neal}, \citenamefont {B.},
  \citenamefont {Zrinzo}, \citenamefont {Hariz}, \citenamefont {Foltynie},
  \citenamefont {Limousin}, \citenamefont {Ashkan}, \citenamefont {FitzGerald},
  \citenamefont {Green}, \citenamefont {Aziz},\ and\ \citenamefont
  {Brown}}]{Little-13}%
  \BibitemOpen
  \bibfield  {author} {\bibinfo {author} {\bibfnamefont {S.}~\bibnamefont
  {Little}}, \bibinfo {author} {\bibfnamefont {A.}~\bibnamefont {Pogosyan}},
  \bibinfo {author} {\bibfnamefont {S.}~\bibnamefont {Neal}}, \bibinfo {author}
  {\bibfnamefont {Z.}~\bibnamefont {B.}}, \bibinfo {author} {\bibfnamefont
  {L.}~\bibnamefont {Zrinzo}}, \bibinfo {author} {\bibfnamefont
  {M.}~\bibnamefont {Hariz}}, \bibinfo {author} {\bibfnamefont
  {T.}~\bibnamefont {Foltynie}}, \bibinfo {author} {\bibfnamefont
  {P.}~\bibnamefont {Limousin}}, \bibinfo {author} {\bibfnamefont
  {K.}~\bibnamefont {Ashkan}}, \bibinfo {author} {\bibfnamefont
  {J.}~\bibnamefont {FitzGerald}}, \bibinfo {author} {\bibfnamefont
  {A.}~\bibnamefont {Green}}, \bibinfo {author} {\bibfnamefont
  {T.}~\bibnamefont {Aziz}}, \ and\ \bibinfo {author} {\bibfnamefont
  {P.}~\bibnamefont {Brown}},\ }\bibfield  {title} {\enquote {\bibinfo {title}
  {Adaptive deep brain stimulation in advanced {P}arkinson disease},}\
  }\href@noop {} {\bibfield  {journal} {\bibinfo  {journal} {Ann Neurol.}\
  }\textbf {\bibinfo {volume} {74}},\ \bibinfo {pages} {449} (\bibinfo {year}
  {2013})}\BibitemShut {NoStop}%
\bibitem [{\citenamefont {Herzog}, \citenamefont {W\"org\"otter},\ and\
  \citenamefont {Parlitz}(2018)}]{Parlitz-18}%
  \BibitemOpen
  \bibfield  {author} {\bibinfo {author} {\bibfnamefont {S.}~\bibnamefont
  {Herzog}}, \bibinfo {author} {\bibfnamefont {F.}~\bibnamefont
  {W\"org\"otter}}, \ and\ \bibinfo {author} {\bibfnamefont {U.}~\bibnamefont
  {Parlitz}},\ }\bibfield  {title} {\enquote {\bibinfo {title} {Data-driven
  modeling and prediction of complex spatio-temporal dynamics in excitable
  media},}\ }\href@noop {} {\bibfield  {journal} {\bibinfo  {journal}
  {Frontiers in Applied Mathematics and Statistics}\ }\textbf {\bibinfo
  {volume} {4}},\ \bibinfo {pages} {60} (\bibinfo {year} {2018})}\BibitemShut
  {NoStop}%
\bibitem [{\citenamefont {Pathak}\ \emph {et~al.}(2018)\citenamefont {Pathak},
  \citenamefont {Hunt}, \citenamefont {Girvan}, \citenamefont {Lu},\ and\
  \citenamefont {Ott}}]{PhysRevLett.120.024102}%
  \BibitemOpen
  \bibfield  {author} {\bibinfo {author} {\bibfnamefont {J.}~\bibnamefont
  {Pathak}}, \bibinfo {author} {\bibfnamefont {B.}~\bibnamefont {Hunt}},
  \bibinfo {author} {\bibfnamefont {M.}~\bibnamefont {Girvan}}, \bibinfo
  {author} {\bibfnamefont {Z.}~\bibnamefont {Lu}}, \ and\ \bibinfo {author}
  {\bibfnamefont {E.}~\bibnamefont {Ott}},\ }\bibfield  {title} {\enquote
  {\bibinfo {title} {Model-free prediction of large spatiotemporally chaotic
  systems from data: {A} reservoir computing approach},}\ }\href@noop {}
  {\bibfield  {journal} {\bibinfo  {journal} {Phys. Rev. Lett.}\ }\textbf
  {\bibinfo {volume} {120}},\ \bibinfo {pages} {024102} (\bibinfo {year}
  {2018})}\BibitemShut {NoStop}%
\bibitem [{\citenamefont {Zimmermann}\ and\ \citenamefont
  {Parlitz}(2018)}]{Zimmermann-Parlitz-18}%
  \BibitemOpen
  \bibfield  {author} {\bibinfo {author} {\bibfnamefont {R.~S.}\ \bibnamefont
  {Zimmermann}}\ and\ \bibinfo {author} {\bibfnamefont {U.}~\bibnamefont
  {Parlitz}},\ }\bibfield  {title} {\enquote {\bibinfo {title} {Observing
  spatio-temporal dynamics of excitable media using reservoir computing},}\
  }\href@noop {} {\bibfield  {journal} {\bibinfo  {journal} {Chaos}\ }\textbf
  {\bibinfo {volume} {28}},\ \bibinfo {pages} {043118} (\bibinfo {year}
  {2018})}\BibitemShut {NoStop}%
\bibitem [{\citenamefont {Quade}\ \emph {et~al.}(2018)\citenamefont {Quade},
  \citenamefont {Abel}, \citenamefont {{Nathan Kutz}},\ and\ \citenamefont
  {Brunton}}]{Quade-18}%
  \BibitemOpen
  \bibfield  {author} {\bibinfo {author} {\bibfnamefont {M.}~\bibnamefont
  {Quade}}, \bibinfo {author} {\bibfnamefont {M.}~\bibnamefont {Abel}},
  \bibinfo {author} {\bibfnamefont {J.}~\bibnamefont {{Nathan Kutz}}}, \ and\
  \bibinfo {author} {\bibfnamefont {S.~L.}\ \bibnamefont {Brunton}},\
  }\bibfield  {title} {\enquote {\bibinfo {title} {Sparse identification of
  nonlinear dynamics for rapid model recovery},}\ }\href@noop {} {\bibfield
  {journal} {\bibinfo  {journal} {Chaos}\ }\textbf {\bibinfo {volume} {28}},\
  \bibinfo {pages} {063116} (\bibinfo {year} {2018})}\BibitemShut {NoStop}%
\bibitem [{\citenamefont {Cestnik}\ and\ \citenamefont
  {Abel}(2019)}]{Cestnik-Abel-19}%
  \BibitemOpen
  \bibfield  {author} {\bibinfo {author} {\bibfnamefont {R.}~\bibnamefont
  {Cestnik}}\ and\ \bibinfo {author} {\bibfnamefont {M.}~\bibnamefont {Abel}},\
  }\bibfield  {title} {\enquote {\bibinfo {title} {Inferring the dynamics of
  oscillatory systems using recurrent neural networks},}\ }\href@noop {}
  {\bibfield  {journal} {\bibinfo  {journal} {Chaos}\ }\textbf {\bibinfo
  {volume} {29}},\ \bibinfo {pages} {063128} (\bibinfo {year}
  {2019})}\BibitemShut {NoStop}%
\bibitem [{\citenamefont {Weng}\ \emph {et~al.}(2019)\citenamefont {Weng},
  \citenamefont {Yang}, \citenamefont {Gu}, \citenamefont {Zhang},\ and\
  \citenamefont {Small}}]{PhysRevE.99.042203}%
  \BibitemOpen
  \bibfield  {author} {\bibinfo {author} {\bibfnamefont {T.}~\bibnamefont
  {Weng}}, \bibinfo {author} {\bibfnamefont {H.}~\bibnamefont {Yang}}, \bibinfo
  {author} {\bibfnamefont {C.}~\bibnamefont {Gu}}, \bibinfo {author}
  {\bibfnamefont {J.}~\bibnamefont {Zhang}}, \ and\ \bibinfo {author}
  {\bibfnamefont {M.}~\bibnamefont {Small}},\ }\bibfield  {title} {\enquote
  {\bibinfo {title} {Synchronization of chaotic systems and their
  machine-learning models},}\ }\href@noop {} {\bibfield  {journal} {\bibinfo
  {journal} {Phys. Rev. E}\ }\textbf {\bibinfo {volume} {99}},\ \bibinfo
  {pages} {042203} (\bibinfo {year} {2019})}\BibitemShut {NoStop}%
\bibitem [{\citenamefont {Yeo}\ and\ \citenamefont
  {Melnyk}(2019)}]{Yeo-Melnyk-19}%
  \BibitemOpen
  \bibfield  {author} {\bibinfo {author} {\bibfnamefont {K.}~\bibnamefont
  {Yeo}}\ and\ \bibinfo {author} {\bibfnamefont {I.}~\bibnamefont {Melnyk}},\
  }\bibfield  {title} {\enquote {\bibinfo {title} {Deep learning algorithm for
  data-driven simulation of noisy dynamical system},}\ }\href@noop {}
  {\bibfield  {journal} {\bibinfo  {journal} {J. of Computational Physics}\
  }\textbf {\bibinfo {volume} {376}},\ \bibinfo {pages} {1212--1231} (\bibinfo
  {year} {2019})}\BibitemShut {NoStop}%
\bibitem [{\citenamefont {Sutton}\ and\ \citenamefont
  {Barto}(2018)}]{RL-Sutton1998}%
  \BibitemOpen
  \bibfield  {author} {\bibinfo {author} {\bibfnamefont {R.~S.}\ \bibnamefont
  {Sutton}}\ and\ \bibinfo {author} {\bibfnamefont {A.~G.}\ \bibnamefont
  {Barto}},\ }\href {http://incompleteideas.net/book/the-book-2nd.html} {\emph
  {\bibinfo {title} {Reinforcement Learning: An Introduction}}},\ \bibinfo
  {edition} {2nd}\ ed.\ (\bibinfo  {publisher} {The MIT Press},\ \bibinfo
  {year} {2018})\BibitemShut {NoStop}%
\bibitem [{\citenamefont {Kaelbling}, \citenamefont {Littman},\ and\
  \citenamefont {Moore}(1996)}]{RL-surveymain-kaelbling1996reinforcement}%
  \BibitemOpen
  \bibfield  {author} {\bibinfo {author} {\bibfnamefont {L.~P.}\ \bibnamefont
  {Kaelbling}}, \bibinfo {author} {\bibfnamefont {M.~L.}\ \bibnamefont
  {Littman}}, \ and\ \bibinfo {author} {\bibfnamefont {A.~W.}\ \bibnamefont
  {Moore}},\ }\bibfield  {title} {\enquote {\bibinfo {title} {Reinforcement
  learning: A survey},}\ }\href@noop {} {\bibfield  {journal} {\bibinfo
  {journal} {Journal of artificial intelligence research}\ }\textbf {\bibinfo
  {volume} {4}},\ \bibinfo {pages} {237--285} (\bibinfo {year}
  {1996})}\BibitemShut {NoStop}%
\bibitem [{\citenamefont {Schulman}\ \emph {et~al.}(2017)\citenamefont
  {Schulman}, \citenamefont {Wolski}, \citenamefont {Dhariwal}, \citenamefont
  {Radford},\ and\ \citenamefont {Klimov}}]{RL-PPO-schulman2017}%
  \BibitemOpen
  \bibfield  {author} {\bibinfo {author} {\bibfnamefont {J.}~\bibnamefont
  {Schulman}}, \bibinfo {author} {\bibfnamefont {F.}~\bibnamefont {Wolski}},
  \bibinfo {author} {\bibfnamefont {P.}~\bibnamefont {Dhariwal}}, \bibinfo
  {author} {\bibfnamefont {A.}~\bibnamefont {Radford}}, \ and\ \bibinfo
  {author} {\bibfnamefont {O.}~\bibnamefont {Klimov}},\ }\bibfield  {title}
  {\enquote {\bibinfo {title} {Proximal policy optimization algorithms},}\
  }\href {http://arxiv.org/abs/1707.06347} {\bibfield  {journal} {\bibinfo
  {journal} {CoRR}\ }\textbf {\bibinfo {volume} {abs/1707.06347}} (\bibinfo
  {year} {2017})},\ \Eprint {http://arxiv.org/abs/1707.06347}
  {arXiv:1707.06347} \BibitemShut {NoStop}%
\bibitem [{\citenamefont {Hill}\ \emph {et~al.}(2018)\citenamefont {Hill},
  \citenamefont {Raffin}, \citenamefont {Ernestus}, \citenamefont {Gleave},
  \citenamefont {Traore}, \citenamefont {Dhariwal}, \citenamefont {Hesse},
  \citenamefont {Klimov}, \citenamefont {Nichol}, \citenamefont {Plappert},
  \citenamefont {Radford}, \citenamefont {Schulman}, \citenamefont {Sidor},\
  and\ \citenamefont {Wu}}]{RL-stable-baselines}%
  \BibitemOpen
  \bibfield  {author} {\bibinfo {author} {\bibfnamefont {A.}~\bibnamefont
  {Hill}}, \bibinfo {author} {\bibfnamefont {A.}~\bibnamefont {Raffin}},
  \bibinfo {author} {\bibfnamefont {M.}~\bibnamefont {Ernestus}}, \bibinfo
  {author} {\bibfnamefont {A.}~\bibnamefont {Gleave}}, \bibinfo {author}
  {\bibfnamefont {R.}~\bibnamefont {Traore}}, \bibinfo {author} {\bibfnamefont
  {P.}~\bibnamefont {Dhariwal}}, \bibinfo {author} {\bibfnamefont
  {C.}~\bibnamefont {Hesse}}, \bibinfo {author} {\bibfnamefont
  {O.}~\bibnamefont {Klimov}}, \bibinfo {author} {\bibfnamefont
  {A.}~\bibnamefont {Nichol}}, \bibinfo {author} {\bibfnamefont
  {M.}~\bibnamefont {Plappert}}, \bibinfo {author} {\bibfnamefont
  {A.}~\bibnamefont {Radford}}, \bibinfo {author} {\bibfnamefont
  {J.}~\bibnamefont {Schulman}}, \bibinfo {author} {\bibfnamefont
  {S.}~\bibnamefont {Sidor}}, \ and\ \bibinfo {author} {\bibfnamefont
  {Y.}~\bibnamefont {Wu}},\ }\href@noop {} {\enquote {\bibinfo {title} {Stable
  baselines},}\ }\bibinfo {howpublished}
  {\url{https://github.com/hill-a/stable-baselines}} (\bibinfo {year}
  {2018})\BibitemShut {NoStop}%
\bibitem [{\citenamefont {Hindmarsh}\ and\ \citenamefont
  {Rose}(1984)}]{Hindmarsh-Rose-84}%
  \BibitemOpen
  \bibfield  {author} {\bibinfo {author} {\bibfnamefont {J.~L.}\ \bibnamefont
  {Hindmarsh}}\ and\ \bibinfo {author} {\bibfnamefont {R.~M.}\ \bibnamefont
  {Rose}},\ }\bibfield  {title} {\enquote {\bibinfo {title} {A model for
  neuronal bursting using three coupled first order differential equations},}\
  }\href@noop {} {\bibfield  {journal} {\bibinfo  {journal} {Proc. Roy. Soc.
  London Ser. B}\ }\textbf {\bibinfo {volume} {221}},\ \bibinfo {pages} {87}
  (\bibinfo {year} {1984})}\BibitemShut {NoStop}%
\bibitem [{Note1()}]{Note1}%
  \BibitemOpen
  \bibinfo {note} {The complete cycle of the diagram in Fig.~\ref {fig:diagram}
  is a multiple of $\delta $.}\BibitemShut {Stop}%
\bibitem [{Note2()}]{Note2}%
  \BibitemOpen
  \bibinfo {note} {The currently available DBS devices deliver a pulsatile
  stimulation with a frequency of about 120 Hz. In particular, the pulse shape
  shown in Fig.~\ref {fig:pulseform}(b) is used.}\BibitemShut {Stop}%
\bibitem [{\citenamefont {Ng}, \citenamefont {Harada},\ and\ \citenamefont
  {Russell}(1999)}]{RL-rewardshaping}%
  \BibitemOpen
  \bibfield  {author} {\bibinfo {author} {\bibfnamefont {A.~Y.}\ \bibnamefont
  {Ng}}, \bibinfo {author} {\bibfnamefont {D.}~\bibnamefont {Harada}}, \ and\
  \bibinfo {author} {\bibfnamefont {S.}~\bibnamefont {Russell}},\ }\bibfield
  {title} {\enquote {\bibinfo {title} {Policy invariance under reward
  transformations: Theory and application to reward shaping},}\ }in\ \href@noop
  {} {\emph {\bibinfo {booktitle} {Proceedings of the Sixteenth International
  Conference on Machine Learning}}}\ (\bibinfo {organization} {ICML},\ \bibinfo
  {year} {1999})\ p.\ \bibinfo {pages} {278–287}\BibitemShut {NoStop}%
\bibitem [{Note3()}]{Note3}%
  \BibitemOpen
  \bibinfo {note} {$T$ is a macroscopic time-scale equal to the duration of
  stimuli application. For prospective wearable DBS systems, $T$ could be
  considered infinitely large.}\BibitemShut {Stop}%
\bibitem [{\citenamefont {Sutton}\ \emph {et~al.}(1999)\citenamefont {Sutton},
  \citenamefont {McAllester}, \citenamefont {Singh},\ and\ \citenamefont
  {Mansour}}]{RL-PolicyTheorem}%
  \BibitemOpen
  \bibfield  {author} {\bibinfo {author} {\bibfnamefont {R.~S.}\ \bibnamefont
  {Sutton}}, \bibinfo {author} {\bibfnamefont {D.}~\bibnamefont {McAllester}},
  \bibinfo {author} {\bibfnamefont {S.}~\bibnamefont {Singh}}, \ and\ \bibinfo
  {author} {\bibfnamefont {Y.}~\bibnamefont {Mansour}},\ }\bibfield  {title}
  {\enquote {\bibinfo {title} {Policy gradient methods for reinforcement
  learning with function approximation},}\ }\href@noop {} {\bibfield  {journal}
  {\bibinfo  {journal} {Neural Information Processing Systems}\ }\textbf
  {\bibinfo {volume} {12}},\ \bibinfo {pages} {1057–1063} (\bibinfo {year}
  {1999})}\BibitemShut {NoStop}%
\bibitem [{\citenamefont {Abadi}\ \emph {et~al.}(2015)\citenamefont {Abadi},
  \citenamefont {Agarwal}, \citenamefont {Barham}, \citenamefont {Brevdo},
  \citenamefont {Chen}, \citenamefont {Citro}, \citenamefont {Corrado},
  \citenamefont {Davis}, \citenamefont {Dean}, \citenamefont {Devin},
  \citenamefont {Ghemawat}, \citenamefont {Goodfellow}, \citenamefont {Harp},
  \citenamefont {Irving}, \citenamefont {Isard}, \citenamefont {Jia},
  \citenamefont {Jozefowicz}, \citenamefont {Kaiser}, \citenamefont {Kudlur},
  \citenamefont {Levenberg}, \citenamefont {Man\'{e}}, \citenamefont {Monga},
  \citenamefont {Moore}, \citenamefont {Murray}, \citenamefont {Olah},
  \citenamefont {Schuster}, \citenamefont {Shlens}, \citenamefont {Steiner},
  \citenamefont {Sutskever}, \citenamefont {Talwar}, \citenamefont {Tucker},
  \citenamefont {Vanhoucke}, \citenamefont {Vasudevan}, \citenamefont
  {Vi\'{e}gas}, \citenamefont {Vinyals}, \citenamefont {Warden}, \citenamefont
  {Wattenberg}, \citenamefont {Wicke}, \citenamefont {Yu},\ and\ \citenamefont
  {Zheng}}]{tensorflow2015-whitepaper}%
  \BibitemOpen
  \bibfield  {author} {\bibinfo {author} {\bibfnamefont {M.}~\bibnamefont
  {Abadi}}, \bibinfo {author} {\bibfnamefont {A.}~\bibnamefont {Agarwal}},
  \bibinfo {author} {\bibfnamefont {P.}~\bibnamefont {Barham}}, \bibinfo
  {author} {\bibfnamefont {E.}~\bibnamefont {Brevdo}}, \bibinfo {author}
  {\bibfnamefont {Z.}~\bibnamefont {Chen}}, \bibinfo {author} {\bibfnamefont
  {C.}~\bibnamefont {Citro}}, \bibinfo {author} {\bibfnamefont {G.~S.}\
  \bibnamefont {Corrado}}, \bibinfo {author} {\bibfnamefont {A.}~\bibnamefont
  {Davis}}, \bibinfo {author} {\bibfnamefont {J.}~\bibnamefont {Dean}},
  \bibinfo {author} {\bibfnamefont {M.}~\bibnamefont {Devin}}, \bibinfo
  {author} {\bibfnamefont {S.}~\bibnamefont {Ghemawat}}, \bibinfo {author}
  {\bibfnamefont {I.}~\bibnamefont {Goodfellow}}, \bibinfo {author}
  {\bibfnamefont {A.}~\bibnamefont {Harp}}, \bibinfo {author} {\bibfnamefont
  {G.}~\bibnamefont {Irving}}, \bibinfo {author} {\bibfnamefont
  {M.}~\bibnamefont {Isard}}, \bibinfo {author} {\bibfnamefont
  {Y.}~\bibnamefont {Jia}}, \bibinfo {author} {\bibfnamefont {R.}~\bibnamefont
  {Jozefowicz}}, \bibinfo {author} {\bibfnamefont {L.}~\bibnamefont {Kaiser}},
  \bibinfo {author} {\bibfnamefont {M.}~\bibnamefont {Kudlur}}, \bibinfo
  {author} {\bibfnamefont {J.}~\bibnamefont {Levenberg}}, \bibinfo {author}
  {\bibfnamefont {D.}~\bibnamefont {Man\'{e}}}, \bibinfo {author}
  {\bibfnamefont {R.}~\bibnamefont {Monga}}, \bibinfo {author} {\bibfnamefont
  {S.}~\bibnamefont {Moore}}, \bibinfo {author} {\bibfnamefont
  {D.}~\bibnamefont {Murray}}, \bibinfo {author} {\bibfnamefont
  {C.}~\bibnamefont {Olah}}, \bibinfo {author} {\bibfnamefont {M.}~\bibnamefont
  {Schuster}}, \bibinfo {author} {\bibfnamefont {J.}~\bibnamefont {Shlens}},
  \bibinfo {author} {\bibfnamefont {B.}~\bibnamefont {Steiner}}, \bibinfo
  {author} {\bibfnamefont {I.}~\bibnamefont {Sutskever}}, \bibinfo {author}
  {\bibfnamefont {K.}~\bibnamefont {Talwar}}, \bibinfo {author} {\bibfnamefont
  {P.}~\bibnamefont {Tucker}}, \bibinfo {author} {\bibfnamefont
  {V.}~\bibnamefont {Vanhoucke}}, \bibinfo {author} {\bibfnamefont
  {V.}~\bibnamefont {Vasudevan}}, \bibinfo {author} {\bibfnamefont
  {F.}~\bibnamefont {Vi\'{e}gas}}, \bibinfo {author} {\bibfnamefont
  {O.}~\bibnamefont {Vinyals}}, \bibinfo {author} {\bibfnamefont
  {P.}~\bibnamefont {Warden}}, \bibinfo {author} {\bibfnamefont
  {M.}~\bibnamefont {Wattenberg}}, \bibinfo {author} {\bibfnamefont
  {M.}~\bibnamefont {Wicke}}, \bibinfo {author} {\bibfnamefont
  {Y.}~\bibnamefont {Yu}}, \ and\ \bibinfo {author} {\bibfnamefont
  {X.}~\bibnamefont {Zheng}},\ }\href {http://tensorflow.org/} {\enquote
  {\bibinfo {title} {{TensorFlow}: Large-scale machine learning on
  heterogeneous systems},}\ } (\bibinfo {year} {2015}),\ \bibinfo {note}
  {software available from tensorflow.org}\BibitemShut {NoStop}%
\bibitem [{\citenamefont {Kuramoto}(1984)}]{Kuramoto-84}%
  \BibitemOpen
  \bibfield  {author} {\bibinfo {author} {\bibfnamefont {Y.}~\bibnamefont
  {Kuramoto}},\ }\href@noop {} {\emph {\bibinfo {title} {Chemical Oscillations,
  Waves and Turbulence}}}\ (\bibinfo  {publisher} {Springer},\ \bibinfo
  {address} {Berlin},\ \bibinfo {year} {1984})\BibitemShut {NoStop}%
\bibitem [{\citenamefont {Pikovsky}\ and\ \citenamefont
  {Ruffo}(1999)}]{Pikovsky-Ruffo-99}%
  \BibitemOpen
  \bibfield  {author} {\bibinfo {author} {\bibfnamefont {A.}~\bibnamefont
  {Pikovsky}}\ and\ \bibinfo {author} {\bibfnamefont {S.}~\bibnamefont
  {Ruffo}},\ }\bibfield  {title} {\enquote {\bibinfo {title} {Finite-size
  effects in a population of interacting oscillators},}\ }\href@noop {}
  {\bibfield  {journal} {\bibinfo  {journal} {Phys. Rev. E}\ }\textbf {\bibinfo
  {volume} {59}},\ \bibinfo {pages} {1633--1636} (\bibinfo {year}
  {1999})}\BibitemShut {NoStop}%
\bibitem [{Note4()}]{Note4}%
  \BibitemOpen
  \bibinfo {note} {The level of the finite-size fluctuations can be estimated
  as $\protect \text {std}(X)$ for the sub-threshold values of the coupling
  $\varepsilon $.}\BibitemShut {Stop}%
\bibitem [{\citenamefont {Arulkumaran}\ \emph {et~al.}(2017)\citenamefont
  {Arulkumaran}, \citenamefont {Deisenroth}, \citenamefont {Brundage},\ and\
  \citenamefont {Bharath}}]{RL-Deep-arulkumaran2017}%
  \BibitemOpen
  \bibfield  {author} {\bibinfo {author} {\bibfnamefont {K.}~\bibnamefont
  {Arulkumaran}}, \bibinfo {author} {\bibfnamefont {M.~P.}\ \bibnamefont
  {Deisenroth}}, \bibinfo {author} {\bibfnamefont {M.}~\bibnamefont
  {Brundage}}, \ and\ \bibinfo {author} {\bibfnamefont {A.~A.}\ \bibnamefont
  {Bharath}},\ }\bibfield  {title} {\enquote {\bibinfo {title} {Deep
  reinforcement learning: A brief survey},}\ }\href@noop {} {\bibfield
  {journal} {\bibinfo  {journal} {IEEE Signal Processing Magazine}\ }\textbf
  {\bibinfo {volume} {34}},\ \bibinfo {pages} {26--38} (\bibinfo {year}
  {2017})}\BibitemShut {NoStop}%
\bibitem [{\citenamefont {Dylov}, \citenamefont {Waller},\ and\ \citenamefont
  {Fleischer}(2011{\natexlab{a}})}]{dylov2011nonlinear}%
  \BibitemOpen
  \bibfield  {author} {\bibinfo {author} {\bibfnamefont {D.~V.}\ \bibnamefont
  {Dylov}}, \bibinfo {author} {\bibfnamefont {L.}~\bibnamefont {Waller}}, \
  and\ \bibinfo {author} {\bibfnamefont {J.~W.}\ \bibnamefont {Fleischer}},\
  }\bibfield  {title} {\enquote {\bibinfo {title} {Nonlinear restoration of
  diffused images via seeded instability},}\ }\href@noop {} {\bibfield
  {journal} {\bibinfo  {journal} {IEEE Journal of Selected Topics in Quantum
  Electronics}\ }\textbf {\bibinfo {volume} {18}},\ \bibinfo {pages} {916--925}
  (\bibinfo {year} {2011}{\natexlab{a}})}\BibitemShut {NoStop}%
\bibitem [{\citenamefont {Dylov}, \citenamefont {Waller},\ and\ \citenamefont
  {Fleischer}(2011{\natexlab{b}})}]{dylov2011instability}%
  \BibitemOpen
  \bibfield  {author} {\bibinfo {author} {\bibfnamefont {D.~V.}\ \bibnamefont
  {Dylov}}, \bibinfo {author} {\bibfnamefont {L.}~\bibnamefont {Waller}}, \
  and\ \bibinfo {author} {\bibfnamefont {J.~W.}\ \bibnamefont {Fleischer}},\
  }\bibfield  {title} {\enquote {\bibinfo {title} {Instability-driven recovery
  of diffused images},}\ }\href@noop {} {\bibfield  {journal} {\bibinfo
  {journal} {Optics letters}\ }\textbf {\bibinfo {volume} {36}},\ \bibinfo
  {pages} {3711--3713} (\bibinfo {year} {2011}{\natexlab{b}})}\BibitemShut
  {NoStop}%
\bibitem [{\citenamefont {Dylov}\ and\ \citenamefont
  {Fleischer}(2010)}]{dylov2010natphotSR}%
  \BibitemOpen
  \bibfield  {author} {\bibinfo {author} {\bibfnamefont {D.~V.}\ \bibnamefont
  {Dylov}}\ and\ \bibinfo {author} {\bibfnamefont {J.~W.}\ \bibnamefont
  {Fleischer}},\ }\bibfield  {title} {\enquote {\bibinfo {title} {Nonlinear
  self-filtering of noisy images via dynamical stochastic resonance},}\
  }\href@noop {} {\bibfield  {journal} {\bibinfo  {journal} {Nature Photonics}\
  }\textbf {\bibinfo {volume} {4}},\ \bibinfo {pages} {323} (\bibinfo {year}
  {2010})}\BibitemShut {NoStop}%
\bibitem [{\citenamefont {Busoniu}, \citenamefont {Babuska},\ and\
  \citenamefont {De~Schutter}(2006)}]{RL-Multiagent-busoniu2006multi}%
  \BibitemOpen
  \bibfield  {author} {\bibinfo {author} {\bibfnamefont {L.}~\bibnamefont
  {Busoniu}}, \bibinfo {author} {\bibfnamefont {R.}~\bibnamefont {Babuska}}, \
  and\ \bibinfo {author} {\bibfnamefont {B.}~\bibnamefont {De~Schutter}},\
  }\bibfield  {title} {\enquote {\bibinfo {title} {Multi-agent reinforcement
  learning: A survey},}\ }in\ \href@noop {} {\emph {\bibinfo {booktitle} {2006
  9th International Conference on Control, Automation, Robotics and Vision}}}\
  (\bibinfo {organization} {IEEE},\ \bibinfo {year} {2006})\ pp.\ \bibinfo
  {pages} {1--6}\BibitemShut {NoStop}%
\bibitem [{Note5()}]{Note5}%
  \BibitemOpen
  \bibinfo {note} {Notice that feedback-based techniques reported in [\cite
  {Popovych_et_al-17}] also do not use any information about the model of the
  system but explicitly assume that pathological activity emerges due to
  synchrony.}\BibitemShut {Stop}%
\bibitem [{\citenamefont {Sanz~Leon}\ \emph {et~al.}(2013)\citenamefont
  {Sanz~Leon}, \citenamefont {Knock}, \citenamefont {Woodman}, \citenamefont
  {Domide}, \citenamefont {Mersmann}, \citenamefont {McIntosh},\ and\
  \citenamefont {Jirsa}}]{TheVirtualBrain}%
  \BibitemOpen
  \bibfield  {author} {\bibinfo {author} {\bibfnamefont {P.}~\bibnamefont
  {Sanz~Leon}}, \bibinfo {author} {\bibfnamefont {S.}~\bibnamefont {Knock}},
  \bibinfo {author} {\bibfnamefont {M.}~\bibnamefont {Woodman}}, \bibinfo
  {author} {\bibfnamefont {L.}~\bibnamefont {Domide}}, \bibinfo {author}
  {\bibfnamefont {J.}~\bibnamefont {Mersmann}}, \bibinfo {author}
  {\bibfnamefont {A.}~\bibnamefont {McIntosh}}, \ and\ \bibinfo {author}
  {\bibfnamefont {V.}~\bibnamefont {Jirsa}},\ }\bibfield  {title} {\enquote
  {\bibinfo {title} {The virtual brain: a simulator of primate brain network
  dynamics},}\ }\href {\doibase 10.3389/fninf.2013.00010} {\bibfield  {journal}
  {\bibinfo  {journal} {Frontiers in Neuroinformatics}\ }\textbf {\bibinfo
  {volume} {7}},\ \bibinfo {pages} {10} (\bibinfo {year} {2013})}\BibitemShut
  {NoStop}%
\bibitem [{\citenamefont {Tass}\ and\ \citenamefont
  {Majtanik}(2006)}]{Tass-Majtanik-06}%
  \BibitemOpen
  \bibfield  {author} {\bibinfo {author} {\bibfnamefont {P.~A.}\ \bibnamefont
  {Tass}}\ and\ \bibinfo {author} {\bibfnamefont {M.}~\bibnamefont
  {Majtanik}},\ }\bibfield  {title} {\enquote {\bibinfo {title} {Long-term
  anti-kindling effects of desynchronizing brain stimulation: a theoretical
  study},}\ }\href@noop {} {\bibfield  {journal} {\bibinfo  {journal}
  {Biological cybernetics}\ }\textbf {\bibinfo {volume} {94}},\ \bibinfo
  {pages} {58--66} (\bibinfo {year} {2006})}\BibitemShut {NoStop}%
\bibitem [{\citenamefont {Blandini}\ and\ \citenamefont
  {Armentero}(2012)}]{ParkinsonAnimalModel}%
  \BibitemOpen
  \bibfield  {author} {\bibinfo {author} {\bibfnamefont {F.}~\bibnamefont
  {Blandini}}\ and\ \bibinfo {author} {\bibfnamefont {M.-T.}\ \bibnamefont
  {Armentero}},\ }\bibfield  {title} {\enquote {\bibinfo {title} {Animal models
  of {P}arkinson’s disease},}\ }\href {\doibase
  10.1111/j.1742-4658.2012.08491.x} {\bibfield  {journal} {\bibinfo  {journal}
  {The FEBS Journal}\ }\textbf {\bibinfo {volume} {279}},\ \bibinfo {pages}
  {1156--1166} (\bibinfo {year} {2012})},\ \Eprint
  {http://arxiv.org/abs/https://febs.onlinelibrary.wiley.com/doi/pdf/10.1111/j.1742-4658.2012.08491.x}
  {https://febs.onlinelibrary.wiley.com/doi/pdf/10.1111/j.1742-4658.2012.08491.x}
  \BibitemShut {NoStop}%
\bibitem [{Note6()}]{Note6}%
  \BibitemOpen
  \bibinfo {note} {Notice that the previously developed feedback-based
  suppression techniques were analyzed theoretically only for the unrealistic
  continuous-time stimulation and their modification to the pulsatile
  stimulation was an \protect \emph {ad hoc}, relying on the assumption of a
  smooth continuous envelope.}\BibitemShut {Stop}%
\end{thebibliography}
%

\end{document}